\documentclass[12pt,a4paper,final]{iopart}

\usepackage{iopams}  
\usepackage{graphicx}
\usepackage[breaklinks=true,colorlinks=true,linkcolor=blue,urlcolor=blue,citecolor=blue]{hyperref}

\usepackage{braket}
\expandafter\let\csname equation*\endcsname\relax
\expandafter\let\csname endequation*\endcsname\relax
\usepackage{amsmath}
\usepackage{float}
\usepackage[toc,page]{appendix}
\usepackage{varwidth}
\allowdisplaybreaks

\begin{document}
\title[The Dynamics of Entropies at the Onset of Interactions]{The Dynamics of Entropies at the Onset of Interactions}

\author{Emily Kendall$^{1,2,3}$ and Achim Kempf$^{1,2}$}
\address{$^1$Perimeter Institute for Theoretical Physics, 31 Caroline St N, Waterloo, Ontario, N2L 2Y5, Canada}
\address{$^2$Departments of Applied Mathematics and Physics, and Waterloo Centre for Astrophysics, University of Waterloo, Waterloo, Ontario, N2L 3G1, Canada}
\address{$^3$Department of Physics, The University of Auckland, Private Bag 92019, Auckland, New Zealand}
\ead{eken000@aucklanduni.ac.nz and akempf@uwaterloo.ca}

\begin{abstract}
At the onset of an interaction between two initially independent systems, each system tends to experience an increase in its $n$-R\'enyi entropies, such as its von Neumann entropy ($n=1$) and its mixedness ($n=2$). 
We here ask which properties of a system determine how quickly its R\'enyi entropies increase and, therefore, how sensitive the system is to becoming entangled.  
We find that the rate at which the $n$-R\'enyi entropy increases in an interaction is determined by a quantity which we term the $n$-fragility of the system. 
The $2$-fragility is closely related to the notion of $2$-norm coherence, in that it too quantifies the extent to which a density matrix is `off-diagonal' with respect to the eigenbasis of a reference operator.
Nevertheless, the 2-fragility is not a coherence monotone in the resource theoretic sense since it depends also on the eigenvalues of the reference operator. It is this additional sensitivity to the eigenvalues of the reference operator, here the interaction Hamiltonian, which enables the 2-fragility to quantify the rate of entropy production in interactions. 
We give an example using the light-matter interaction and we anticipate applications to the study of the rates at which two systems exchange classical and quantum information when starting to interact.

\end{abstract}

\vspace{2pc}

\section{Introduction}\label{sec:intro}

At the heart of emerging quantum technologies, such as quantum computing, quantum communication, quantum cryptography and quantum sensing, see, e.g., \cite{Gisin2002, Riedel2015, Boyer2017, Grimm2019, Zhang2019, Lasota2019, Klco2019}, are the phenomena of entanglement and coherent superposition. Both entanglement and coherent superposition, or `coherence' for short, are purely quantum phenomena for which there exist no classical analogues. Both can be viewed as physical resources which can undergo complex dynamics and much work is presently being done to develop their formal resource theories, see, e.g., \cite{Streltsov:2015xia,Coecke2016,Streltsov2017,Contreras2019}. 
The study of entanglement and coherence is important also beyond quantum technologies, for example, in cosmology where the quantum fluctuations of the metric and of the inflaton field in the very early universe decohered and then seeded cosmic structure formation, see, e.g., \cite{Kiefer2012}.

In the present paper, we investigate the interplay of coherence and entanglement at the onset of interactions. Our aim is to
develop tools for studying the onset of interactions information theoretically. The new tools are designed to help answer, for example, the question of what determines at what rate two systems will hand classical and quantum information to each other when starting to interact. To this end, we here introduce the new notion of $n$-fragility.  
We define the $n$-fragility as that quantity which determines the speed with which a system's $n$-R\'enyi entropy increases at the onset of an interaction. 
The $n$-fragility, as a new tool for tracking the dynamics of entropies at the onset of interactions, should then be useful, for example, to study at what speed classical and quantum channel capacities are being established at the onset of an interaction.

The new notion of $n$-fragility is closely related to measures of coherence but differs in an important way.   
For an intuitive picture, let us consider a cat in coherent superposition of being at two  locations that are well-separated (i.e., without overlap). Any coherence monotone of this state, such as the  $2$-norm coherence, is
independent of how close or far apart these two locations are. However, the speed with which the cat's state tends to decohere due to interactions does not only depend on the amount of coherence of the state. The speed of decoherence also grows with the distance between the cat's two locations.

The new notion of $n$-fragility that we introduce here is designed to quantify this `fragility' of the purity of a system under interaction. In the example of the cat, the $n$-fragility must, therefore, be sensitive to the distance between the two virtual locations of the cat. For example, it is the $2$-fragility which quantifies the rate at which such a cat state increases its mixedness at the onset of an interaction. 

Let us recall that coherence monotones in the resource-theoretic sense are dependent on a choice of reference basis, such as here the eigenbasis of an operator in an interaction Hamiltonian. The $n$-fragility will have to be sensitive not only to the eigenbasis but also to the eigenvalues of that reference operator. This is because the eigenvalues of the reference operator, which is here part of the interaction Hamiltonian, do influence how fast the interaction and therefore the decoherence proceeds in the respective Hilbert space dimensions. 

This means that the $n$-fragility is not a coherence monotone in the resource-theoretic sense: operations that merely swap basis vectors of the reference basis are incoherent and by definition they are unable, therefore, to increase the value of what is called a coherence monotone. The $n$-fragility, however, can increase under these operations because they also involve a swap of the eigenvalues. The only exception is the special case of qubits, where there are only two eigenvalues and therefore only one spacing. The $2$-fragility then reduces to the $2$-norm coherence up to a prefactor, the prefactor being given by the spacing of the two eigenvalues of the reference operator. 

Further, as we will show, the full picture is that, at the onset of an interaction between two systems, A and B, with interaction Hamiltonian $H_A\otimes H_B$, the leading order change in the $n$-R{\'e}nyi entropy of system A is the second order and the full expression is given by the product of two factors. One is the $n$-fragility of system A with respect to the reference operator $H_A$ and the other factor is the variance $(\Delta H_B)^2$ of system B. We also find that in the limit of a pure state, the $2$-fragility reduces to the variance.

While the $1$- and $2$-R\'enyi entropies tend to be the most commonly occurring, we here cover all $n$-R\'enyi entropies. This is because covering all $n$-R\'enyi entropies involves little additional calculational overhead and they can be useful, for example, in the Replica trick \cite{Rangamani:2016dms}.

We begin with explicit examples of simple qubit systems in order to gain intuition before generalising to generic systems. Finally, we apply our results to the example of an idealised model of the light-matter interaction, the Jaynes-Cummings model.

\section{Coherence versus entanglement in simple systems}

\subsection{Coherent and incoherent uncertainty in qubit systems}\label{sec:2-dim}

For a system in a state $\rho$, the uncertainty, $\Delta B$, in an observable, $B$, generally possesses coherent as well as incoherent contributions. The incoherent contribution to the uncertainty $\Delta B$  is due to classical ignorance or the `mixedness' of the state. Here, the mixedness, $\mu$, is defined through $\mu:=1-\gamma$ where $\gamma := \Tr(\rho^2)$ is the purity. 

In contrast, the coherent contribution to $\Delta B$ originates in quantum fluctuations that arise when the state $\rho$ possesses nonzero off-diagonal elements in the eigenbasis of the observable $B$. This suggests that in order to describe the magnitude of the coherent contribution to $\Delta B$, we use the  so-called  2-norm coherence, $c_{2,B}(\rho)$, which is obtained\footnote{The 2-norm coherence has a particularly simple interpretation for pure states, $\rho=\ket{\psi}\bra{\psi}$, in which case $c_{2,B}(\rho)$ is a measure of `cattiness' in the sense that it characterises the extent to which the state resembles an equally weighted coherent superposition of the eigenstates of $B$.} by summing up the squared moduli of the off-diagonal elements, see \cite{coherencemeas}: 
\begin{equation}\label{eq:coherence-measure}
c_{2,B}(\rho):= \vert\vert (\mathbb{I}-\mathcal{D}_B)\rho \ \vert\vert_2^2 =\sum_{i\neq j=1}^d \vert \rho_{ij}\vert^2,
\end{equation}
Here,  $\vert\vert X\vert\vert_2:=\sqrt{Tr(X^\dagger X)}$ is the Hilbert-Schmidt 2-norm and $\mathcal{D}_B$ is the dephasing superoperator which sets the off-diagonal elements of $\rho$ in the eigenbasis of $B$ to zero. 

For qubit systems, i.e., for two-level systems, there exists a particularly simple decomposition of the variance $(\Delta B)^2$ into the sum of a coherent and an incoherent contribution in terms of $c_{2B}(\rho)$ and $\mu$. 
To see this, let us denote the eigenvectors and eigenvalues of $B$ by $\ket{b_x}$, $\ket{b_y}$ and $b_x$,  $b_y$ respectively. For a generic state, $\rho$, of the qubit, writing $\rho_{xx}:=\bra{b_x}\rho\ket{b_x}$ etc, we have:
\begin{align}\label{eq:2d-purity}
    \gamma &= \Tr(\rho^2) = \rho_{xx}^2+\rho_{yy}^2+2\vert\rho_{xy}\vert^2,
\\ \label{eq:2d-coherence}
    c_{2,B} &= \vert\rho_{xy}\vert^2+\vert\rho_{yx}\vert^2=2\vert\rho_{xy}\vert^2.
\end{align}
Using
\begin{equation}\label{eq:2d-uncertainty}
    (\Delta B)^2=\Tr(\rho B^2)-\Big(\Tr(\rho B)\Big)^2=(\rho_{xx}-\rho_{xx}^2)(b_x-b_y)^2, 
\end{equation}
we now obtain
\begin{equation}\label{eq:2-dim-variance}
    (\Delta B)^2=\frac{1-\gamma+c_{2b}}{2}(b_x-b_y)^2.
\end{equation} 
For given values of $b_x$ and $b_y$, the maximum achievable uncertainty\footnote{Note that both the maximally coherent pure state and the maximally mixed state have maximal uncertainty. Indeed, any combination of $c_{2b}$ and $\gamma$ which satisfies $c_{2b}-\gamma=-1/2$ (or, equivalently, $\rho_{xx}=1/2$), will correspond to a state with maximal uncertainty.} is given by $(\Delta B)^2_{max} = \frac{1}{4}(b_x-b_y)^2
$. We can therefore reformulate Equation \ref{eq:2-dim-variance} to obtain: 
\begin{equation}\label{eq:fractional-variance}
    \frac{(\Delta B)^2}{2(\Delta B)^2_{max}}= c_{2B} + \mu.
\end{equation}
Equation \ref{eq:fractional-variance} is a decomposition of the normalised variance of a qubit system into the sum of a coherent contribution $c_{2,B}$ and an incoherent contribution $\mu$. On one hand, $\mu$ can be viewed as the contribution to the variance that is due to classical ignorance. On the other hand, any mixedness, $\mu$, can also always be viewed, via suitably dilating the system, as arising from entanglement with an ancilla. Therefore, we can also view Equation \ref{eq:fractional-variance} as a decomposition of $(\Delta B)^2$ into a contribution from coherence and a contribution from entanglement. 
We will make use of this decomposition in later sections.

\subsection{Unitary evolution of coherence and entanglement}\label{sec:dynamics}

Let us now briefly review the evolution of the two contributions to the variance, coherence and entanglement (or equivalently, coherence and mixedness) as our qubit system interacts with a qubit ancilla. We consider two cases of interactions between the system and the ancilla. In the first case, the interaction is described by time evolution operators of the form $U(t)=e^{it A\otimes B}$, where $A$ is a self-adjoint operator acting on the ancillary system and where 
$B$ is the observable of the system that we considered above. 
The second type of interactions between a system and an ancilla that we will consider is described by time evolution operators of the form $U(t)=e^{i t A\otimes H_B}$, where $H_B$ is an operator which does \it not \rm commute with the observable $B$. In both cases, we here for simplicity assume that the interaction Hamiltonian is dominant during the interaction, i.e., that we can neglect the free Hamiltonians of the system and ancilla during the interaction. 

\subsubsection{Case 1: Interactions in which the creation of entanglement and of coherence are antagonistic.}\label{sec:antagonistic-action}

In the case of the first type of interaction, described by $U(t)=e^{it A\otimes B}$, 
entanglement and coherence behave antagonistically. Namely, whenever the entanglement between the system and the ancilla increases and whenever, therefore, the mixedness $\mu$ increases, the 2-norm coherence of the system decreases, and vice versa. To see this, we notice that $[1\otimes B,U(t)]=0$ and that, therefore, $d\Delta B(t)/dt=0$. In Equation \ref{eq:fractional-variance}, the left hand side is, therefore, constant, implying that indeed any drop in the coherence $c_{2,B}$ is accompanied by a corresponding rise in the mixedness $\mu$ and a corresponding rise in the entanglement with the ancilla, and vice versa.

Of course, this type of interaction can also be interpreted as a measurement in the sense that the ancilla system gains information about the  observable $B$. This is because in $U(t)=e^{it A\otimes B}$ the operator $B$ controls the extent to which $A$ moves a pointer variable, $P$, of the ancilla system during the time evolution. Here, a pointer variable $P$ of the ancilla is any observable of the ancilla system which does not commute with $A$ and which, therefore, experiences a nontrivial time evolution under $U(t)$. When we view the ancilla as a measurement device, the time-evolution of $\langle P\rangle$ is the movement of the pointer.  

If, for example, the system is in an eigenstate $\vert b\rangle$ of $B$ with eigenvalue $b$, then the evolution of the ancilla is governed by the time evolution operator $e^{itbA}$. Let us then consider any observable, $P$, of the ancilla that does not commute with $A$. The amount by which the expectation value $\langle P\rangle$ moves during the interaction is determined by the value $b$. Therefore, $P$ can be used as a pointer variable for a measurement of $B$. Generally, for any initial state of our system, the ancilla system is in this sense `measuring' $B$. 
Concretely, for
\begin{equation}\label{eq:entangling_unitary}
    U(t)=e^{i\varepsilon t \sigma_x\otimes B},
\end{equation}
a generic initial state that is pure and unentangled reads  
\begin{equation}\label{eq:psi}
\ket{\psi_0}=\frac{\left(\ket{x^+}+r\ket{x^-}\right)\otimes\left(\ket{b_x}+s\ket{b_y}\right)}{\sqrt{\left(1+\vert r\vert^2\right)\left(1+\vert s\vert^2\right)}},
\end{equation}
where $\vert x^+\rangle$ and $\ket{x^-}$ are the eigenstates of $\sigma_x$. The time-dependent mixedness and 2-norm coherence of our system are easily found to be:
\begin{equation}\label{eq:purity}
    \mu(t) =1-\frac{1}{(1+\vert s\vert^2)^2}\Bigg(1+\vert s\vert^4+2\vert s\vert^2\frac{1+\vert r\vert^4+2\vert r\vert^2\cos\big[2\varepsilon t(b_x-b_y)\big]}{(1+\vert r\vert^2)^2}\Bigg)
\end{equation}
\begin{equation}\label{eq:coherence-simple}
    c_{2,B}\big(\rho_B(t)\big)=\frac{2\vert s\vert^2\big(1+\vert r\vert^4+2\vert r\vert^2\cos[2\varepsilon t(b_x-b_y)]\big)}{(1+\vert r\vert^2)^2(1+\vert s\vert^2)^2}
\end{equation}
Using Equation \ref{eq:fractional-variance}, their sum yields the variance: 
\begin{equation}
    \big(\Delta B\big)^2=\Bigg(\frac{\vert s\vert\big(b_x-b_y\big)}{1+\vert s\vert^2}\Bigg)^2
\end{equation}
We notice that the variance is constant in time, as is expected due to $[1\otimes B,U(t)]=0$. 
Since the sum of the mixedness $\mu(t)$ and the 2-norm coherence $c_{2B}(t)$ is constant, we see that whenever the interaction creates entanglement between the system and the ancilla and whenever, therefore, the interaction increases the mixedness of the system, the $2$-norm coherence of the system correspondingly diminishes.
This is of course the phenomenon that the system `decoheres' through entanglement creation, in the sense that the reduced state $\rho$ of the system literally loses (2-norm) coherence, which means that its off-diagonal elements in the eigenbasis of $B$ diminish.

\subsubsection{Case 2: Evolution in which the creation of entanglement and of coherence are not antagonistic.}\label{genamp}

Let us now consider the case of the second type of interaction between system and ancilla, described by time evolution operators of the form $U(t)=e^{i t A\otimes H_B}$ where $H_B$ is an operator which does not commute with the observable $B$. In this case, since the uncertainty $\Delta B(t)$ is not conserved, entanglement and coherence need not evolve antagonistically, i.e., entanglement may be generated while the coherence of the system simultaneously increases. 

We can understand this situation in terms of the antagonistic case that we considered in the previous subsection, Section \ref{sec:antagonistic-action}. This is because, from the discussion of Section \ref{sec:antagonistic-action}, we can conclude that an interaction of the type $U(t)=e^{i t A\otimes H_B}$ can be viewed as the ancilla system measuring the system's observable $H_B$. The density matrix of the system, therefore, now loses its off diagonal elements in the eigenbasis of $H_B$ and not in the eigenbasis of $B$. Since the 2-norm coherence $c_{2B}$ is defined relative to the eigenbasis of the observable $B$, and since $H_B$ and $B$ are diagonal in different bases, we have that while the system's coherence with respect to $H_B$ decreases during the interaction, the system's coherence with respect to $B$ need not necessarily decrease and it can, in fact, increase. How this behavior occurs is explicitly illustrated in Appendix \ref{appHb}.

\section{Entanglement generation at the onset of interactions in generic quantum systems: introducing the new notion of n-fragility}\label{extension}

Beyond qubit systems, we will now consider generic systems interacting with a generic ancilla. Generally, the time evolution can then no longer be solved exactly and we will, therefore, pursue a perturbative analysis of the behavior of coherence and entanglement at the onset of the interaction between system and ancilla. 

\subsection{Perturbative analysis of the nonperturbatively known qubit case}

In order to see what type of behavior to expect, let us first consider the perturbative expansion of the exact result that we obtained in the previous section for the interaction of a qubit system with a qubit ancilla described by the unitary $U(t)=e^{it A\otimes B}=e^{i\varepsilon t \sigma_x\otimes B}$. To this end, we Taylor expand the purity $\gamma(t)$ using the exact solution for $\mu(t) = 1-\gamma(t)$ in Equation \ref{eq:purity}. Since we start with a pure state, we have $\gamma(0)=1$.  The first time derivative vanishes, $\gamma'(0)=0$. The Taylor series of $\gamma(t)$ truncated at the second order, $\gamma^{(2)}(t)$, is the lowest nontrivial order. A short calculation yields: 
\begin{equation}\label{eq:2nd_order_uncert}
    \gamma^{(2)}(t)=1-2\varepsilon^2 t^2\big(\Delta\sigma_x\big)^2\big(\Delta B\big)^2.
\end{equation} 
This shows that, when a qubit system and a qubit ancilla are initially pure and unentangled, then 
the initial rate at which the qubit system loses its purity through entanglement creation at the onset of the interaction is proportional to the variance, $(\Delta B)^2$ of the system's operator, $B$, in the interaction Hamiltonian $H_i=A\otimes B$.
Further, since the qubit and the ancilla play  interchangeable roles in the interaction, Equation \ref{eq:purity} is invariant under the exchange of  $\Delta A$ and $\Delta B$. This means that 
$\Delta B$ is also proportional to the rate at which the system decreases the purity of the ancilla. 

We conclude that, if the primary qubit system and its qubit ancilla start out pure and unentangled, then $\Delta B$ is proportional to both the primary system's propensity to reduce the purity of the ancilla and its propensity to reduce its own purity. We will see, however, that for initial states that are mixed, this is generally not the case. This may indeed be anticipated from the fact that a system that is in the maximally mixed state can reduce the purity of an ancilla in an interaction, but it cannot further reduce its own purity. 

Indeed, we will find that the ability of a system to depurify itself at the onset of an interaction is described by a quantity that is an extension of the notion of $2$-norm coherence which we will call the $2$-fragility. More generally, we will obtain a notion of $n$-fragility by considering the procilivity for the system's $n$-R\'enyi entropy to change at the onset of an interaction. 

\subsection{Variance as a determiner of depurification rate}

We now analyse the dynamics of the purity and of R\'enyi entropies at the onset of interactions between generic systems of any finite dimensions that start out in generic mixed initial states.

We begin with the case where the system  starts out in a generic mixed state $\rho_B(t_0)$ and is then coupled to an ancilla that is initially in a pure state $\ket{\phi}$, with the combined system then evolving through the time evolution operator
 $   U(t)=e^{itA\otimes B}$. 
We obtain that
 $   \rho_{tot}(t)=e^{itA\otimes B}\ket{\phi}\bra{\phi}\otimes\rho_B(t_0)e^{-itA\otimes B},$
from which the purity of the ancilla system, $\gamma_A(t)$, is:
 $   \gamma_A(t)=\operatorname{Tr}\bigg[\Big(\operatorname{Tr_B}\big[\rho_{tot}(t)\big]\Big)^2\bigg]$.
We write the time evolution of the purity as a Taylor series in $t$:
As in the case of qubits, we again have that  $\omega_1=0$, so that the leading order is the second order in $t$: 

\begin{equation}
    \gamma_A(t)^{(2)}=1+\frac{1}{2}t^2 \operatorname{Tr}\Bigg[\ \frac{\partial^2}{\partial t^2}\bigg(\Big(\operatorname{Tr_S}\big[\rho_{tot}(t)\big]\Big)^2\bigg) \ \Bigg] \Bigg|_{t=0}
\end{equation}

A short calculation yields: 
\begin{equation}\label{puritygen}
     \gamma_A(t)^{(2)}=1-2t^2\big(\Delta A\big)^2\big(\Delta B\big)^2,
\end{equation}
where 
 $   \big(\Delta A\big)^2=\bra{\phi}A^2\ket{\phi}-\bra{\phi}A\ket{\phi}^2$
and $    \big(\Delta B\big)^2=\operatorname{Tr_S}\big[B^2\rho_B(t_0)\big]-\Big(\operatorname{Tr_S}\big[B\rho_B(t_0)\big]\Big)^2.$
We notice that the perturbative result in Equation \ref{puritygen} has exactly the same form as Equation \ref{eq:2nd_order_uncert}, which we obtained from the exact calculation in the previous simple example. Hence, we see that in general, the leading order contribution to the loss of purity in the ancilla system is proportional to the variance $(\Delta B)^2$. This holds for higher dimensional systems and ancillas, and also when the system starts in a mixed state, as long as the ancilla starts in a pure state. Equation \ref{puritygen} has been previously obtained, for example in \cite{Giulini:1996}. 

Equation \ref{puritygen} shows that the larger the uncertainty $\Delta B$ in the system is, the faster an ancilla that it couples to will lose purity. Interestingly, it does not matter in Equation \ref{puritygen} whether the uncertainty $\Delta B$ is of coherent or incoherent origin. A state which is maximally mixed will have the same proclivity to depurify a fresh ancilla as does a pure, maximally coherent state. While this may seem to be an insensitivity to whether the uncertainty in the system is of quantum or classical origin, we recall that an uncertainty of incoherent origin (or `classical uncertainty'), can of course always be re-interpreted as possessing a quantum origin in the form of entanglement in a suitably dilated system.

Of course, the roles of the system and ancilla can be swapped here by simply replacing their names. We conclude that if the system starts in a pure state and the ancilla in a mixed state then the time evolution of the purity of our system, B, $\big(\gamma_B(t)^{(2)}\big)$, to leading order is again given by Expression \ref{puritygen}. This means, in particular, that if our system B starts out in a pure state, then its proneness for self-depurification in the interaction is proportional to the variance $(\Delta B)^2$. 

\subsection{New result: the $n$-fragility determines a system's rate of self-depurification and $n$-R\'enyi entropy increase at the onset of interactions}\label{int_frag}

We now consider the case where both the system and the ancilla start in a mixed state. At present, we still require the two systems to share no initial entanglement - a criterion which we will later revisit. Namely, we assume that the initial total state is of the form $\rho_A \otimes \rho_B$, and we consider a unitary interaction of the form $e^{itA\otimes B}$. We then define the following quantity:
\begin{equation}
    \gamma_{n,B}(t):=\operatorname{Tr}_{B}\Big[\big(\operatorname{Tr}_A\left[\rho(t)\right]\big)^n\Big]
\end{equation}
Here $n$ is a positive integer. For $n=2$, this quantity is the purity of the state $\rho_B(t)$. For arbitrary $n$, we will refer to this as the $n-$purity. While the $n=2$ case is of obvious significance, we notice that in general, for $n\geq1$, the quantity $\gamma_{n,B}(t)$ is related to the $n$-R\'{e}nyi entropies of the state $\rho_B(t)$, where in the limit $n\rightarrow 1$ we obtain the von Neumann entropy. The $n$-R\'{e}nyi entropies are given by:
\begin{equation}\label{eq:Renyi-state}
  H_n(\rho_B(t)):=\frac{1}{1-n}\operatorname{log}\gamma_{n,B}(t), 
\end{equation}
which may be alternatively expressed as
\begin{equation}\label{eq:Renyi-state-alt}
  H_n(\rho_B(t)):=\frac{1}{1-n}\operatorname{log}\left(\sum_{i}\lambda_i^n(t)\right),
\end{equation}
where the $\lambda_i(t)$ are the eigenvalues of $\rho_B(t)$.

In order to explore the dynamics of the loss of purity and the increase in the $n$-R\'{e}nyi entropies, let us consider the change of the $n$-purity, $\gamma_{n,B}(t)$, over time. To second order in time we have
\begin{equation}
    \gamma_{n,B}(t) = \gamma_{n,B}(0)+\dot{\gamma}_{n,B}(0)t+\frac{1}{2}\ddot{\gamma}_{n,B}(0)t^2+...
\end{equation}
where
\begin{equation}
    \dot{\gamma}_{n,B}(t) = n\operatorname{Tr}_{B}\Big[\Big(\operatorname{Tr}_{A}\big[\rho(t)\big]\Big)^{n-1}\operatorname{Tr}_A\big[\dot{\rho}(t)\big]\Big],
\end{equation}
and
\begin{align}
    \ddot{\gamma}_{n,B}(t) = n\operatorname{Tr}_{B}\Big[&\operatorname{Tr}_A\big[\dot{\rho}(t)\big]\sum_{i=0}^{n-2}\operatorname{Tr}_A\big[\rho(t)\big]^i\operatorname{Tr}_A\big[\dot{\rho}(t)\big]\operatorname{Tr}_A\big[\rho(t)\big]^{n-2-i}\nonumber\\
    &+\operatorname{Tr}_A\big[\rho(t)\big]^{n-1}\operatorname{Tr}_A\big[\ddot{\rho}(t)\big]\Big].
\end{align}
It is easy to verify that at $t=0$ there is no contribution at first order since:
\begin{align}
    \dot{\rho(t)} &= \frac{d}{dt}\Big(e^{it A\otimes B}(\rho_{A,0}\otimes \rho_{B,0}) e^{-it A\otimes B}\Big)\nonumber\\
    &= i\big(A\rho_{A,0}\otimes B\rho_{B,0} - A\rho_{A,0}\otimes B\rho_{B,0}\big)
\end{align}
such that
\begin{align}\label{eq:first-order-zero}
    \dot{\gamma}_{n,B}(0)&=in\operatorname{Tr}_B\Big[\rho_{B,0}^{n-1}\operatorname{Tr}_A\big[B\rho_{B,0}\otimes A\rho_{A,0}-\rho_{B,0}B\otimes \rho_{A,0}A\big]\Big]\nonumber\\
    &=in\operatorname{Tr}_B\Big[\rho_{B,0}^{n-1}\big(B\rho_{B,0}\otimes\operatorname{Tr}_A\big[ A\rho_{A,0}\big]-\rho_{B,0}B\otimes\operatorname{Tr}_A\big[ \rho_{A,0}A\big]\big)\Big]\nonumber\\
    &=in\operatorname{Tr}_B\Big[\rho_{B,0}^{n-1}\operatorname{Tr}_A\big[A\rho_{A,0}\big]\big(B\rho_{B,0}-\rho_{B,0}B\big)\Big]\nonumber\\
    &=0,
\end{align}
where we have twice used the cyclicity of the trace. Hence, to leading (second) order we have:

\begin{align}\label{eq:fragility}
    \ddot{\gamma}_{n,B}(0)&=-n\Big(\operatorname{Tr}_A\big[A\rho_{A,0}\big]\Big)^2\operatorname{Tr}_B\Big[[B,\rho_{B,0}]\sum_{i=0}^{n-2}\rho_{B,0}^{i}[B,\rho_{B,0}]\rho_{B,0}^{n-2-i}\Big]\nonumber\\
    &\qquad +2n\operatorname{Tr}_A\big[A^2\rho_{A,0}\big]\operatorname{Tr}_B\Big[\rho_{B,0}^{n-1}[B,\rho_{B,0}]B\Big]\nonumber\\[1em]
    &=2n\big(\Delta A\big)^2\operatorname{Tr}_B\Big[\rho_{B,0}^{n-1}[B,\rho_{B,0}]B\Big].
\end{align}
We therefore find that in the case of a system starting in a mixed state $\rho_{B,0}$, the reduction of its own $n$-purity at the onset of the interaction with an ancilla is not determined by the variance $(\Delta B)^2$ but instead by the following quantity, which we call the $n$-fragility: 
\begin{equation}
f_n := - \frac{n}{2}\operatorname{Tr}_B\Big[\rho_{B,0}^{n-1}[B,\rho_{B,0}]B\Big]    
\end{equation}
We define the $n$-fragilities with a leading negative constant such that the $n$ fragility is positive, and such that, as we will see later, it reduces to the variance for pure states. While it is clear that the cases $n=2$ and $n = 1$ will have immediate utility in describing the changes in purity and von Neumann entropy in interactions, the complete class of $n-$fragilities will also prove useful in forthcoming work relating to quantum channel capacities, where tools such as the Replica trick, see, e.g., \cite{Rangamani:2016dms}, can be of use. Furthermore, as we will see shortly in our discussion of the 1-fragility, it is useful to begin with the generalised expression for the $n-$fragility, from which we can then consider taking limits (i.e. $n\rightarrow 1$). 

Let us first consider the $n$-fragility for $n=2$, i.e., the tendency of a system in a mixed state to reduce its own $2$-purity, $\gamma_{2,B}(t)$, which is the usual purity. In this case, we have:
\begin{equation}\label{eq:gamma-2-ddot}
     \ddot{\gamma}_{2,B}(0)=2\big(\Delta A\big)^2\operatorname{Tr}_B\Big[\big[B,\rho_{B,0}\big]^2\Big],
\end{equation}
The leading (second) order contribution to the change in the purity of system $B$ is proportional to both the uncertainty in system $A$ and to the 2-fragility, $f_2$:
\begin{equation}\label{eq:2-fragility}
    f_2:=-\frac{1}{2}{Tr}_B\Big[\big[B,\rho_{B,0}\big]^2\Big].
\end{equation}
We arrive here at the finding that it is the $2$-fragility which determines the proclivity of a system to lose its own purity at the onset of the interaction with an ancilla. We also notice already that, as we will discuss in more detail in Sec.\ref{sec3.5}, $f_2$ (and similarly all $f_n$) provide a natural extension of the notion of the quantum coherence of a state $\rho$ with respect to an observable $B$. This is because $-\frac{1}{2}\mbox{Tr}\left([B,\rho]^2\right)$ canonically quantifies the nondiagonality of $\rho$ in the eigenbasis of $B$, namely by measuring the amount of noncommutativity of $\rho$ and $B$.  
For later reference, let us note that we have not assumed that either subsystem is initially pure, nor that the combined system is pure. We have, however, made the assumption that the two subsystems are initially, i.e., at $t=0$, unentangled and that $\rho(t=0)=\rho_{A,0}\otimes\rho_{B,0}$.

Overall, Equation \ref{eq:gamma-2-ddot} tells us that the variance of a system determines its ability to reduce the purity of a system that it interacts with, while the 2-fragility of the system characterises its ability to reduce its own purity. We therefore see that there is generally a difference between the tendency of self-depurification, and the tendency of a system to depurify an ancilla. As we mentioned above, it is intuitive that there has to be a difference between the two proclivities since, for example, in the extreme case of a system in the maximally mixed state, that system can de-purify a system that it interacts with while at the same time its own purity cannot be further reduced.

\subsection{The 2-fragility is bounded by the variance.}

We have seen that the $2$-fragility describes the tendency of a system $S_1$ to reduce its own purity when starting to interact with an ancilla, $S_2$. We also know from Equation \ref{puritygen}, that in the special case where the system $S_1$ starts in a pure state, its tendency to reduce its own purity is determined by the variance $(\Delta S_1)^2$. 
We conclude, therefore, that the 2-fragility must reduce to the variance in the special case where the initial state of the system is pure. Indeed, we can show this directly. At t=0, and for a pure state $\rho_1=\vert\psi_1\rangle\langle\psi_1\vert$:
\begin{equation}
    f_2(S_1)=-\frac{1}{2}\operatorname{Tr}_{1}\Big[\big[S_1,\rho_{1}\big]^2\Big]=-\operatorname{Tr}_1\Big[\rho_1S_1\rho_1S_1-\rho_1S_1^2\Big],
\end{equation}
Here, we used the cyclic property of the trace as well as the fact that $\rho_1^2=\rho_1$ for a pure state. We then express $\rho_1$ as $\vert\psi_1\rangle\langle\psi_1\vert$, and we see that 
\begin{equation}
    f_2(S_1)=-\Big(\langle\psi_1\vert S_1\vert\psi_1\rangle\langle\psi_1\vert S_1\vert\psi_1\rangle-\langle\psi_1\vert S_1^2\vert\psi_1\rangle\Big)=(\Delta S_1)^2,
\end{equation}
as required. Hence we see that the 2-fragility of a \textit{pure} state coincides with the variance of the observable appearing in the interaction Hamiltonian.\footnote{We note that in the limit where the initial state becomes pure, also the $n-$fragility for $n>1$ reduces to the variance.}

Having established that the 2-fragility reduces to the variance for pure states, we now compare the two quantities in the general case of mixed states. The difference between the variance and the $2$-fragility reads:
\begin{equation}\label{eq:frag_var}
    (\Delta H)^2-f_2(H) = \Tr[\rho H^2]-\Tr[\rho H]^2+\frac{1}{2}\Tr\left[[H,\rho]^2\right]
\end{equation}
Let us first re-express this in the eigenbasis of $\rho$, where $\rho_{ij}=p_i\delta_{ij}$ and $0\leq p_i\leq 1$.
\begin{align}
    (\Delta H)^2-f_2(H) &= \sum_{ijk}\rho_{ij}H_{jk} H_{ki}-\sum_{ijkl}\big(\rho_{ij}H_{ji}\rho_{kl} H_{lk} + H_{ij}\rho_{jk}H_{kl}\rho_{li}-\rho_{ij}\rho_{jk} H_{kl}H_{li}\big)\nonumber\\
    &=\sum_{ik}\big(p_i\vert H_{ik}\vert^2-p_i H_{ii} p_k H_{kk}+p_i H_{ik}p_k H_{ki}-p_i^2H_{ik}H_{ki}\big)\nonumber\\
    &=\sum_{ik}\big((\underbrace{p_i-p_i^2}_{\geq 0}+p_i p_k)\vert H_{ik}\vert^2-p_i p_k H_{ii} H_{kk}\big).
\end{align}
Because $0\leq p_i\leq 1$, the difference $p_i-p_i^2 \geq 0$. We will now show that the remainder of the expression is also $\geq 0$, and that, therefore, the variance is always larger than or equal to the 2-fragility. Looking at the last two terms we have:
\begin{equation}\label{eq:positivity}
    \sum_{ik}\big(p_ip_k(\vert H_{ik}\vert^2-H_{ii}H_{kk})\big)=\Tr[\rho H\rho H]-\Tr[\rho H]^2.
\end{equation}
To see that this is $\geq 0$, we first note that
\begin{equation}
    \sum_{ik}p_ip_k\vert H_{ik}\vert^2=\sum_ip_i^2H_{ii}^2+\underbrace{\sum_{i,k\neq i}p_ip_k\vert H_{ik}\vert^2}_{\geq 0},
\end{equation}
Furthermore, we have:
\begin{align}
    &\frac{1}{2}\sum_{ik}(H_{ii}p_i-H_{kk}p_k)^2=\sum_{ik}\big(H_{ii}^2p_i^2-H_{ii}H_{kk}p_ip_k\big)>0\nonumber\\
    &\therefore ~~\sum_{i}H_{ii}^2p_i^2>\sum_{ik}H_{ii}H_{kk}p_ip_k,
\end{align}
and since
\begin{equation}\label{eq:simp_positivity}
    \sum_{ik}p_ip_k\vert H_{ik}\vert^2\geq\sum_{i}H_{ii}^2 p_i^2
\end{equation}
we see that the expressions in Equation \ref{eq:positivity} are positive. Hence, in general, we have that the variance is greater than or equal to the 2-fragility: $(\Delta B)_\rho^2\ge-\frac{1}{2}\mbox{Tr}([\rho,B]^2)$, while the two quantities are equal for pure states. 

We remark here that interactions with interaction Hamiltonians of the form $A \otimes B$ that we consider here are, of course, not always entangling. It is straightforward to see that an example is the case when at least one of the subsystems is in an eigenstate of its part of the interaction Hamiltonian. Both the variance and the fragilities then vanish. The situation becomes more intricate for more general interaction Hamiltonians of the form $A\otimes B + C\otimes D +...$ which we will cover in upcoming work. 

\subsection{The relationship between the 2-fragility and the 2-norm coherence.} \label{sec3.5}
The 2-fragility for generic states, given by the basis-independent expression in Equation \ref{eq:2-fragility}, shows that the 2-fragility is directly a measure of the noncommutativity of the state $\rho$ and the reference operator $B$. Writing the $2$-fragility in the eigenbasis of the operator $B$, the 2-fragility takes a form that is reminiscent of the $2$-norm coherence: 
\begin{equation}\label{generalised-coherence}
    f_2 =\frac{1}{2}\sum_{i, j}(b_i-b_j)^2\vert\rho_{ij}\vert^2.
\end{equation}
We now see that the $2$-fragility is a natural generalization of the notion of $2$-norm coherence, see Equation \ref{eq:coherence-measure}, in that also the $2$-fragility consists of the sum of the moduli squared of the off-diagonal density matrix elements, except that, importantly, the $2$-fragility also takes into account the spacing of the eigenvalues of $B$. 
Only in the special case of qubits, where there is only one eigenvalue spacing, the $2$-fragility coincides exactly with the $2$-norm coherence, up to a prefactor given by the spread of the eigenvalues, $(b_i-b_j)^2$. 

In general, however, the $n$-fragilities do not satisfy the criteria of a coherence monotone. To see this, let us consider a quantum channel which simply swaps basis vectors of the reference operator with respect to which the $2$-norm coherence or $2$-fragility are calculated, i.e., here the operator in the Hamiltonian. This channel is an incoherent operation under which, by definition, coherence monotones cannot increase. The channel will permute
off-diagonal matrix elements within a state's density matrix. Since the magnitudes of the off diagonal elements do not change, the 2-norm coherence will indeed remain the same. However, because of the presence of the $(b_i-b_j)^2$ term in Equation \ref{generalised-coherence}, the associated prefactors of off-diagonal terms in the fragility may differ, and consequently the 2-fragility itself can increase. 

It is this sensitivity of the $n$-fragilities not only to the eigenvectors but also to the eigenvalues of the Hamiltonian which enables the $n$-fragilities to capture the speed at which entropies change at the onset of an interaction. As we will discuss in the Outlook, because the $n$-fragilities capture this speed, they, and related quantities, can be expected to play an important role in the calculation of the speed with which classical and quantum channel capacities arise at the onset of an interaction between two systems.

\subsection{The 1-fragility determines the initial rate of change of the von Neumann entropy}

We can perform a similar analysis to the above, choosing instead the limit $n\to1$ in Equations \ref{eq:Renyi-state} and \ref{eq:Renyi-state-alt}. In this limit we obtain the von Neumann entropy.
We first consider the Taylor expansion of the n'th-R\'{e}nyi entropy:
\begin{equation}
H_{n,B}(t) = H_{n,B}(0) + \dot{H}_{n,B}(0)t + \frac{1}{2}\ddot{H}_{n,B}(0)t^2+...
\end{equation}
As was the case for the purity, it is easy to verify that the first order contribution at t=0 vanishes. Hence, the leading order contribution is given by:
\begin{equation}
    \ddot{H}_{n,B}(0)=\frac{-\dot{\gamma}^2_{n,B}(0)}{(1-n)\gamma^2_{n,B}(0)}+\frac{\ddot{\gamma}_{n,B}(0)}{(1-n)\gamma_{n,B}(0)}
    =\frac{\ddot{\gamma}_{n,B}(0)}{(1-n)\gamma_{n,B}(0)}
\end{equation}
Substituting in the result for $\ddot{\gamma}_{n,B}(0)$ from Equation \ref{eq:fragility} we obtain
\begin{equation}
    \ddot{H}_{n,B}(0)=\frac{2n\big(\Delta A\big)^2\operatorname{Tr}_B\Big[\rho_{B,0}^{n-1}[B,\rho_{B,0}]B\Big]}{(1-n)\operatorname{Tr}_B\big[\rho_{B,0}^{\ n}\big]}.
\end{equation}
We now introduce $\varepsilon=n-1$, and write
\begin{align}
    \ddot{H}_{\varepsilon+1,B}(0)&=\frac{-2(\varepsilon+1)\big(\Delta A\big)^2\operatorname{Tr}_B\Big[\rho_{B,0}^{\varepsilon}[B,\rho_{B,0}]B\Big]}{\varepsilon\operatorname{Tr}_B\big[\rho_{B,0}^{\ \varepsilon+1}\big]}\nonumber\\
    &=\frac{-2(\varepsilon+1)\big(\Delta A\big)^2\operatorname{Tr}_B\Big[\operatorname{exp}\big((\varepsilon\operatorname{log}(\rho_{B,0})\big)[B,\rho_{B,0}]B\Big]}{\varepsilon\operatorname{Tr}_B\big[\rho_{B,0}^{\ \varepsilon+1}\big]}\nonumber\\
     &=\frac{-2(\varepsilon+1)\big(\Delta A\big)^2\operatorname{Tr}_B\Big[\big(1+\varepsilon\operatorname{log}(\rho_{B,0})+\mathcal{O}(\varepsilon^2)\big)[B,\rho_{B,0}]B\Big]}{\varepsilon\operatorname{Tr}_B\big[\rho_{B,0}^{\ \varepsilon+1}\big]}.
\end{align}
Taking the limit that $\varepsilon\rightarrow 0$ (i.e. $n\rightarrow 1$) and keeping only the leading order term, we have 
\begin{equation}
    \underset{n \to 1}{\lim}\ddot{H}_{n,B}(0)=-2\big(\Delta A\big)^2\operatorname{Tr}_B\Big[\operatorname{log}(\rho_{B,0})[B,\rho_{B,0}]B\Big],
\end{equation}
which is the leading order contribution to the change in the von Neumann entropy of the state $\rho_B$, and we may define from this the 1-fragility, 
\begin{equation}\label{testlabel}
    f_{1,B}:=-\operatorname{Tr}_B\Big[\operatorname{log}(\rho_{B,0})[B,\rho_{B,0}]B\Big].
\end{equation}
For completeness, let us consider the special case when the initial state is pure. Working in the eigenbasis of $\rho_{B,0}$, with eigenvalues $\{\lambda_i\}$, we can we can write
\begin{equation}\label{divergence}
    \operatorname{Tr}_B\Big[\operatorname{log}(\rho_{B,0})[B,\rho_{B,0}]B\Big]=\sum_{i\neq k}\big(\lambda_k\operatorname{log}\lambda_i-\lambda_i\operatorname{log}\lambda_i\big)\vert B_{ik}\vert^2,
\end{equation}
where $B_{ik}$ is the $i,k$ matrix element of $B$ in the eigenbasis of $\rho_{B,0}$. In the limit that the state is pure, all eigenvalues go to zero except for one eigenvalue, which goes to 1. Hence, in this limit, the only contribution in the above sum comes from a term $1\operatorname{log}(0)$, so that Equation \ref{testlabel} diverges to $+\infty$ in general. The obvious exception to this arises when $B$ is diagonal in the eigenbasis of the state (i.e. $B_{ik}=0$ for $i\neq k$), in which case the interaction is not entangling.
The two limits of the state becoming pure and of the state starting to commute with the operator $B$, therefore, do not commute. Technically, for pure states, the first time derivative of the von Neumann entropy is zero, while the second diverges and hence, the Taylor expansion is not valid in this special case. Indeed, this is true not just for pure states, but for any state with at least one zero eigenvalue. In practice, states are generically mixed and the Taylor expansion is valid. 

\section{Application to the light-matter interaction}

In practice, any form of communication, including quantum communication, is based on the interaction between light and matter. The development of technologies capable of harnessing the light-matter interaction as a means of quantum information transfer is, therefore, currently a very active area of research, see, e.g., \cite{Northup2014, Patel2016, Lee2015}. 

An idealised description of the light-matter interaction that has proven very useful for the study of the basic principles of relativistic quantum information theory is provided by the Unruh-deWitt model and its variants such as the Jaynes-Cummings model or Quantum Rabi model, see e.g. \cite{Braak2011, Lv:2018cxf, Schneeweiss2018}. 

In the Unruh DeWitt model, the interaction between a first-quantised atom and the second-quantised electromagnetic field is simplified by replacing the electromagnetic field with a massless Klein Gordon field and by restricting attention to only two energy levels of the atom. This projection of the atom's Hilbert space down to two dimensions yields an effective description of the atom as a qubit.

Traditionally, an atom that is simplified in this way is referred to as an Unruh DeWitt (UDW) detector, the term detector referring to  atoms as a detectors of photons through absorption. The UDW detector model was first used to study Hawking radiation and the Unruh effect, see \cite{Birrell1982}. Today, the UDW detector model is in frequent use in the field of relativistic quantum information to study, for example, entanglement extraction from the vacuum, see, e.g., \cite{Valentini1991,Reznik:2002fz,Reznik2005,VerSteeg:2007xs,Martin-Martinez:2013eda}. 

Of particular interest for our purposes here is the fact that the  UDW detector model is useful for the study of the flow of classical and quantum information from atoms, molecules or ions, via field quanta, to other atoms, molecules or ions, \cite{Cliche2010}. We can now compute the $n$-fragilities of the field and the qubit atom, thereby quantifying the proclivities of the atom and the field to increase their $n$-R\'enyi entropies 
at the onset of the interaction. These proclivities are of interest because they should be closely related to the 
susceptibility of the atom and the field to receive quantum information from another, i.e., to hand pre-existing entanglement with an ancilla to another, as the onset of the interaction starts to establish a quantum channel.  

\subsection{Dynamics of the fragilities of a qubit atom interacting with a Klein Gordon field mode}\label{sec:kg_mode}
We consider a field $\phi(x,t)$ in a cavity of side length $\mathrm{L}$. Working in 1+1 dimensions and imposing Dirichlet boundary conditions, we can then write the following mode decomposition:
\begin{align}
    \phi(x,t)&=\sum_{n=1}^{\infty}\sqrt{\frac{1}{\omega_n \mathrm{L}}}\mathrm{sin}\left(\frac{n\pi x}{\mathrm{L}}\right)(a_n^{\dagger}(t)+a_n(t))\nonumber\\
    &= \sum_{n=1}^{\infty}\nu_n(x)(a_n^\dagger (t)+a_n(t))
\end{align}
We take the initial state of the field to be the vacuum state, 
\begin{equation}
    \rho_{f} = \bigotimes_n\ket{0_n}\bra{0_n}.
\end{equation}
We couple this field to a qubit, whose state we expressed in terms of the positive and negative eigenvalue eigenstates of the $\sigma_x$ operator:
\begin{equation}
    \rho_{q}=\alpha\ket{x^+}\bra{x^-}+\alpha^*\ket{x^-}\bra{x^+} + \delta\ket{x^+}\bra{x^+} +(1-\delta)\ket{x^-}\bra{x^-}
\end{equation}

Let us assume that the field and qubit are initially in a product state and unentangled. We couple the qubit to the modes of the field through the Unruh DeWitt interaction Hamiltonian:
\begin{equation}\label{e45}
    H_{\mathrm{int}} = \sigma_{x}\otimes \nu_m(x)(a_m^\dagger+a_m)\bigotimes_{n\neq m}\mathbb{I}_n
\end{equation}
Here, $x$ is the location of the qubit atom. Notice that the standard Unruh DeWitt model assumes that the location of the center of mass of the atom can be described classically. For the inclusion of the quantum properties of the center mass, see \cite{Stritzelberger:2019gnc}.  

In Equation \ref{e45}, we adopted the single-mode approximation, i.e., we assume that the qubit's energy gap is in resonance with a particular cavity mode $m$ that dominates the interaction.
This means that we will now ignore all field modes $n\neq m$ and will drop the subscript $m$, i.e., we will write $a,a^\dagger,\nu,\omega$ for $a_m,a_m^\dagger,\nu_m,\omega_m$. For simplicity, we will also crudely assume that the interaction Hamiltonian dominates, so that we can neglect the free evolution of the field and qubit, i.e., we will assume the unitary time evolution of field and atom to be given by
\begin{equation}\label{eq:field-matter-int}
    U(t)=\exp{[it \nu\sigma_x\otimes(a^\dagger+a)]}.
\end{equation}

We now calculate the fragilities of the field as a function of $t$. 
To do so it will be useful to normal order $(a+a^\dagger)^n$. 
With the definition
\begin{equation}
    \Omega_n = \sum_{i=0}^n {n \choose i} a^{\dagger n-i}a^i.
\end{equation}
we obtain for $n>1$:

\begin{equation}\label{ordering}
    (a+a^\dagger)^n = \begin{cases}
      \Omega_n  +\sum\limits_{i=1}^{n/2} \ \prod\limits_{k=0}^{i-1}\big(2 k + 1\big)\displaystyle\binom{n}{2i}\Omega_{n-2i}, & \  n \ \text{even}, \\[15pt]
      \Omega_n+ \sum\limits_{i=1}^\frac{n-1}{2} \ \prod\limits_{k=0}^{i-1}\big(2k+1\big)\displaystyle{n \choose 2i}\Omega_{n-2i},& \ n \ \text{odd}.
    \end{cases}
\end{equation}

We can now compute the following exact expression for the 2-fragility of the qubit atom under time evolution:

\begin{align}
    &f_{2,q}=4\vert\alpha\vert^2\Big\vert\Tr_{f}[e^{2 i t \nu (a+a^\dagger)}\ket{0}\bra{0}]\Big\vert^2,
\end{align}  
For the $2$-fragility of the field we obtain:
\begin{align}\label{eq:full_field_frag}
    f_{2,f}&=-\frac{1}{2}\Tr\Bigg(2\big(\delta^2 +(1-\delta)^2\big)(a+a^\dagger)\ket{0}\bra{0}\big[(a+a^\dagger),\ket{0}\bra{0}\big]\nonumber\\
    &+\delta(1-\delta)\Big(e^{2it(a+a^\dagger)}\ket{0}\bra{0}e^{-2it(a+a^\dagger)}+e^{-2it(a+a^\dagger)}\ket{0}\bra{0}e^{2it(a+a^\dagger)}\Big)\times\nonumber\\
    &\Big[\big[(a+a^\dagger),\ket{0}\bra{0}\big],(a+a^\dagger)\Big]
    \Bigg)
\end{align}
It is straightforward to verify that for $t=0$ these expressions reduce to the familiar form of the 2-fragility as presented in Equation \ref{eq:2-fragility}. We can now calculate the field's $2$-fragility normal ordering from Equation \ref{ordering}. First, computing the commutators in Equation \ref{eq:full_field_frag} we obtain:
\begin{equation}
    f_{2,f}= \delta^2 + (1-\delta)^2 + \delta(1-\delta)\Tr\big(X\big),
\end{equation}
where,
\begin{align}
    X &= \Big(e^{2it(a+a^\dagger)}\ket{0}\bra{0}e^{-2it(a+a^\dagger)}+e^{-2it(a+a^\dagger)}\ket{0}\bra{0}e^{2it(a+a^\dagger)}\Big)\times\nonumber\\
    & \ \ \ \ \ \Big(\ket{0}\bra{0}+1/\sqrt{2}\big(\ket{0}\bra{2}+\ket{2}\bra{0}\big)-\ket{1}\bra{1}\Big).
\end{align}
Hence, we see that in order to compute the trace, we need only consider the terms in the exponential series which result in $\ket{0}$, $\ket{1}$ or $\ket{2}$ states. We note that for the left action of  $\exp{[\pm it\nu (a+a^\dagger)]}$ on $\ket{0}$ we have:
\begin{align}\label{eq:expansion_1}
    &\exp{[\pm it\nu(a+a^\dagger)]}\ket{0} =\ket{0}+(\pm it\nu)\ket{1}+\nonumber\\[1em]
    &\sum_{n \ \text{even} \ \geq 2}^\infty\frac{(\pm it \nu)^n}{n!}\bigg(\sqrt{n!}\ket{n}+\sum_{i=1}^{n/2}\prod_{k=0}^{i-1}(2 k +1){n \choose 2i}\sqrt{(n-2i)!}\ket{n-2i}\bigg)+\nonumber\\
    &\sum_{n \ \text{odd} \ \geq 3}^\infty\frac{(\pm it\nu)^n}{n!}\bigg(\sqrt{n!}\ket{n}+\sum_{i=1}^{(n-1)/2}\prod_{k=0}^{i-1}(2 k +1){n \choose 2i}\sqrt{(n-2i)!}\ket{n-2i}\bigg).
\end{align}
Likewise for the right-action of the exponential on $\bra{0}$, with the kets replaced with bras in the above.
After some simplification, we obtain an exact expression for the field fragility, namely
\begin{equation}
    f_{2,f} = \delta^2 + (1-\delta)^2 + 2\delta(1-\delta)\big(1-(6+\sqrt{2})t^2\big)e^{-4t^2},
\end{equation}
where we have set $\nu=1$ for simplicity. We plot the behaviour of the $2$-fragility over time for varying values of $\delta$. This is shown in Figure \ref{fig:exact_frag_field}, with further plots given in Appendix \ref{app:lm-plots}.

\begin{figure}[!htb]
    \centering
    \includegraphics[scale = 0.7, trim={0cm 0cm 0cm 0cm}]{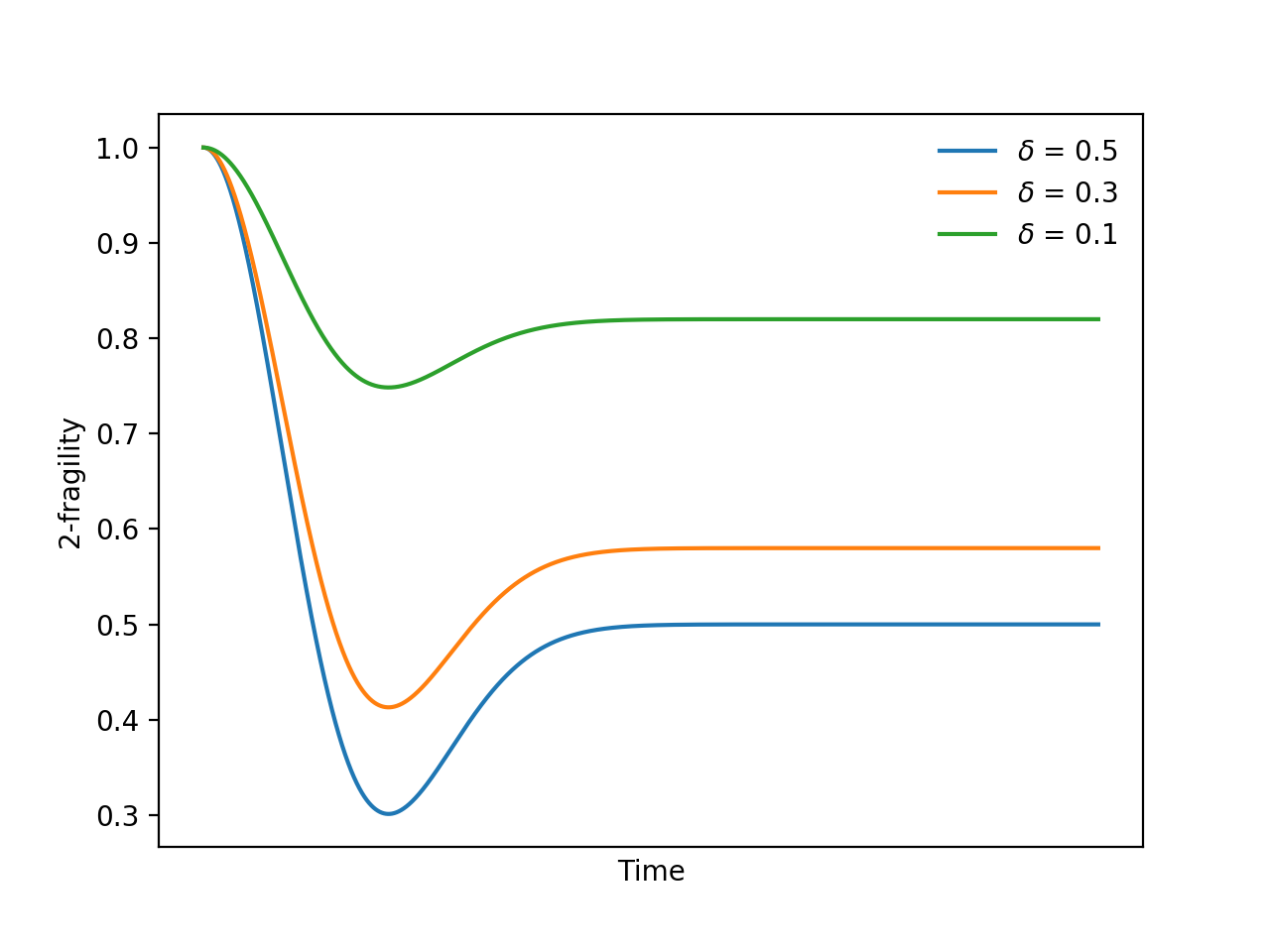}
    \caption{The $2$-fragility of the field as a function of time (arbitrary scale) for the interaction described by Equation \ref{e45}. As the interaction progresses, the fragility reaches a constant value.}    \label{fig:exact_frag_field}
\end{figure}

\subsection{Interpretation}
When we defined the $2$-fragility in  Section \ref{extension}, we showed that the $2$-fragility of a system possesses (at least) two interpretations: 
\begin{enumerate}
\item The $2$-fragility is the procilivity of the system to lose its purity at the onset of an interaction. 
\item The $2$-fragility is a natural extension of the $2$-norm coherence. Namely, the $2$-fragility quantifies the degree of non-commutativity between the state and the observable. 
\end{enumerate}
In the previous subsection, Section \ref{sec:kg_mode}, we calculated the fragilities of the qubit atom and the field mode both at the onset of their interaction and also during their interaction.  

At the onset of their interaction, both interpretations (i) and (ii) apply. A finite time into the interaction between the atom and the field, when they have already become entangled, it would appear that only interpretation (ii) still applies. In fact, however, the $2$-fragilities that we obtain for the field and the atom at finite time still also carry the interpretation (i), namely as the extent to which these systems to lose purity at the onset of interactions that each may incur with a \it new \rm and initially unentangled ancilla system. For example, the $2$-fragility of the field at a finite time, $t$, can be interpreted as the leading order loss of purity of the field at the onset of its interaction with an atom that it had not previously been coupled to.  

\subsection{The case of the field initialised in a thermal state}
\begin{figure}[H]
\centering
\begin{tabular}{cc}
\includegraphics[scale = 0.5, trim={0cm 0cm 0cm 0cm}]{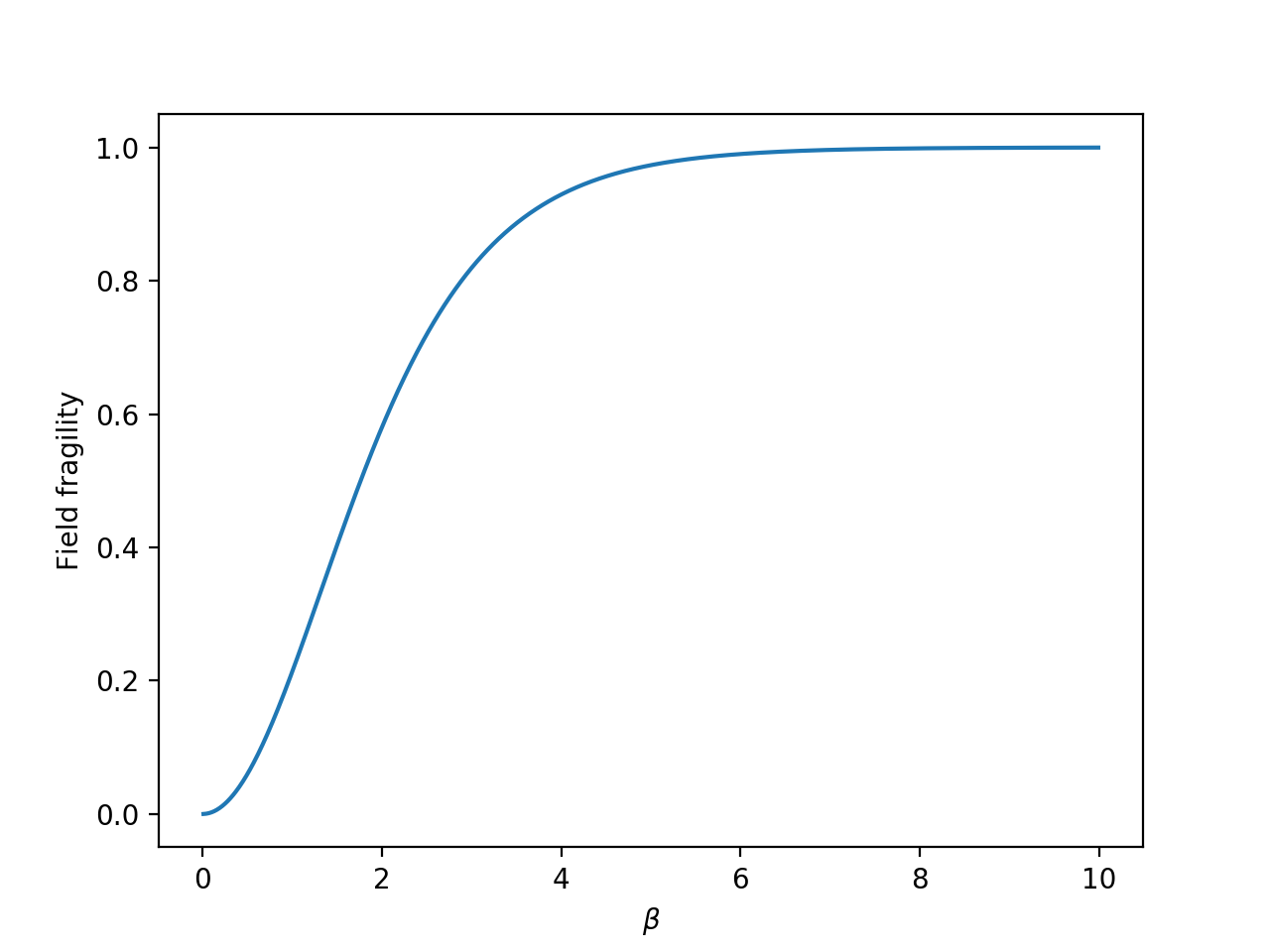} &
{\includegraphics[scale = 0.5, trim={1.5cm 0cm 0cm 0cm}]{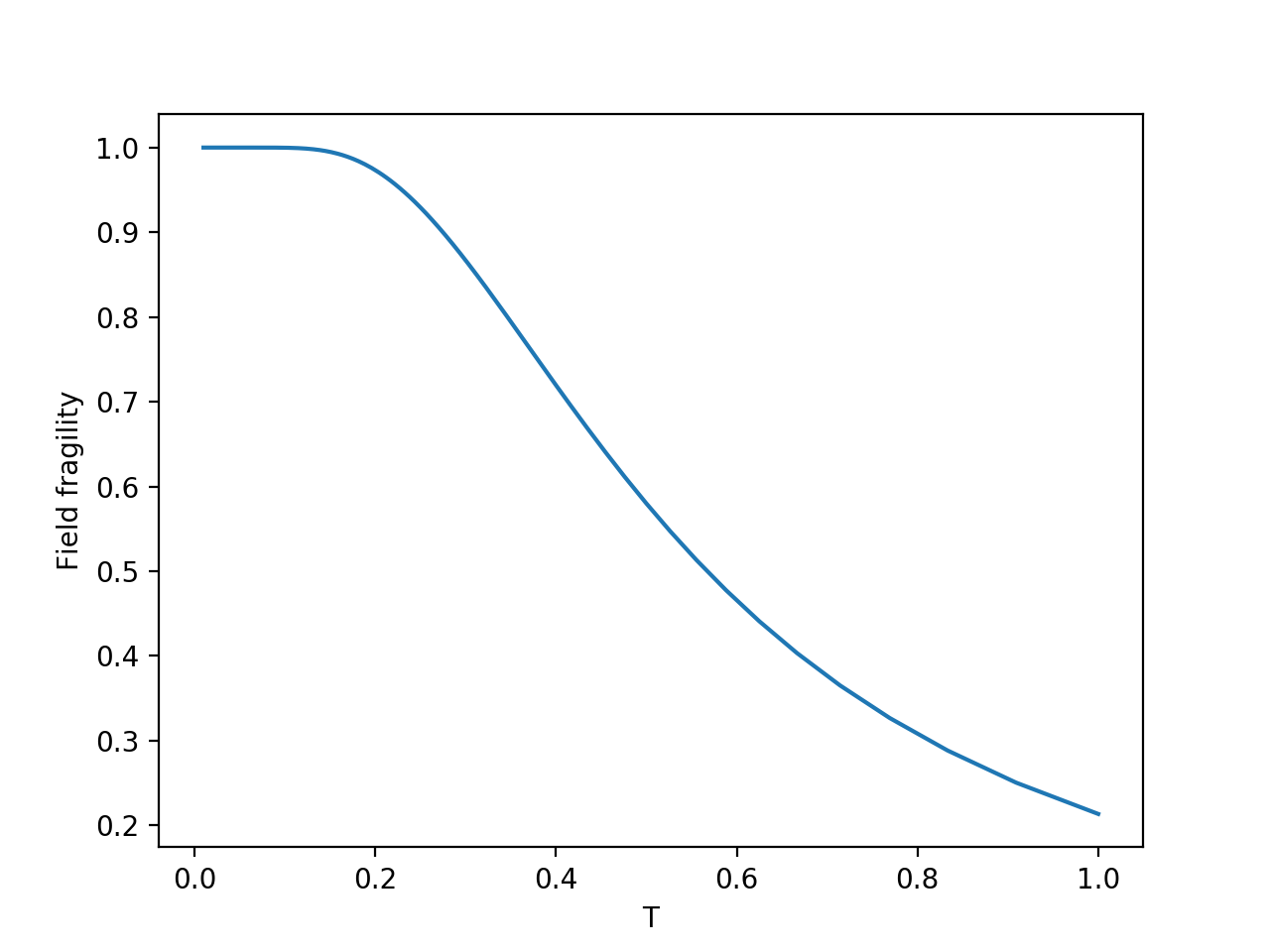}}
\end{tabular}
\caption{The $2$-fragility of the field in a thermal state,  as a function of $\beta$ (left) and $\mathrm{T}$ (right). The $2$-fragility approaches unity as $\beta$ is increased, i.e., it is maximal when the field is in the pure vacuum state.}\label{fig:thermal}
\end{figure}

Let us now calculate the initial $2$-fragility of a field that is in a thermal state at the onset of the interaction. 
In this case, we have for the $2$-fragility of the field:
\begin{equation}\label{eq:thermal}
    f_{2,f} = -\nu^2~\frac{1}{2}\frac{\mathrm{Tr}\bigg(\big[a+a^\dagger,\mathrm{exp}(-\beta \omega a^\dagger a)\big]^2\bigg)}{\mathrm{Tr}\bigg(\mathrm{exp}(-\beta\omega a^\dagger a)\bigg)^2}
\end{equation}
Here, $\beta=1/k\mathrm{T}$ where $k$ is the Boltzmann constant and $\mathrm{T}$ is the temperature. Working in the eigenbasis of the number operator, we obtain:
\begin{equation}
    f_{2,f} =\nu^2 ~\frac{ \sum\limits_{n=0}^\infty(n+1)\Big(\e^{-2\beta\omega(n+1)} + \e^{-2\beta\omega n} - 2\e^{-\beta\omega(n+1)}\e^{-\beta\omega n}\Big)}{\left(\sum\limits_{n=0}^\infty \e^{-\beta\omega n}\right)^2}.
\end{equation}
These geometric series can be summed up to obtain:
\begin{equation}
    f_{2,f} = \nu^2~ \frac{4\sinh^4{[\beta\omega/2]}}{\sinh^2{[\beta\omega]}}
\end{equation}
We show how the $2$-fragility of the field varies with the choice of $\beta$ or, equivalently, $T$, in Figure \ref{fig:thermal}. The plot shows how, as expected, the fragility has its maximum at zero temperature, $\beta\rightarrow\infty$, when the field is in the vacuum state, which is pure. We notice that as the temperature is increased, the fragility initially remains quite flat.  Further, the plot shows how the $2$-fragility eventually decays towards zero as the temperature of the field is increased. This means that as the temperature is assumed larger and larger, the field not only enters the interaction while in a less and less pure state, but the field also exhibits a smaller and smaller tendency to lose its remaining purity in an interaction. 

\section{Conclusions and Outlook}

In addition to entanglement, also coherence can be viewed as a resource for quantum technologies, such as quantum computing, quantum communication and quantum sensing. The intuition behind the concept of coherence is that it is the part of the uncertainty, $\Delta B$, in an observable that is due to fundamental quantum fluctuations instead of being due to classical ignorance, as we illustrated in Section \ref{sec:2-dim}. There, we used as the measure of coherence the $2$-norm coherence which quantifies the off-diagonality of a state $\rho$ in the eigenbasis of a reference observable $B$ by summing up the (squared moduli of the) off-diagonal matrix elements of $\rho$ in the eigenbasis of $B$. 

In this context, we considered a generic system and an ancilla that start in a state $\rho_A\otimes \rho_B$ and that then interact through an interaction Hamiltonian of the form $H=A\otimes B$ with the first and second tensor factors belonging to the ancilla and system respectively. 
Usually, the onset of an interaction generates entanglement which then lowers the purity of both the system and the ancilla.
However, if the system is in a state $\rho_B$ which is diagonal in the eigenbasis of $B$, i.e., if $[\rho_B,B]=0$, then the $2$-norm coherence of the state is zero and the interaction will not reduce the purity. 

This suggests that the rates at which a system loses its purity, gains mixedness and gains general $n$-R\'enyi entropies in an interaction is governed by quantities that are related to the coherence. This is indeed the case. What we established is that it is the $n$-fragilities, 
$f_n = - \frac{n}{2}\operatorname{Tr}_B\Big[\rho_{B}^{n-1}[B,\rho_{B}]B\Big]  
$, which quantify the leading order tendency of a system to increase its mixedness and general $n$-R\'enyi entropies at the onset of an interaction. 

The $n$-fragilities are closely related to measures of coherence in the sense that they too are measures of the noncommutativity of $\rho$ and $B$. We showed also that, in a basis, the $2$-fragility $f_2=-\frac{1}{2}\mbox{Tr}([\rho,B]^2)$ and the $2$-norm coherence are structurally quite similar. However, we also showed that, for an important reason, none of the $n$-fragilities are coherence monotones in the resource-theoretic sense.     
The reason is that coherence monotones of a state can depend only the eigenbasis of the reference observable while the $n$-fragilities must depend on both the eigenbasis as well as on the eigenvalues of the reference observable. The reference observable is here the system's operator in the interaction Hamiltonian. The $n$-fragilities are sensitive to the Hamiltonian's eigenvalues because the $n$-fragilities are designed to quantify the magnitude of the speed with which the interaction  increases the $n$-R\'enyi entropies including the von Neumann entropy $(n=1)$ and the mixedness $(n=2)$.

The new notion of $n$-fragility should be useful, therefore, for example, in the study of the classical and quantum capacities, as measured in terms of entropies, of the quantum channel that arises when two systems start to interact. 
Let us recall that the quantum channel capacity is the ability of a quantum channel to transmit pre-existing entanglement with an ancilla. Since the $n$-fragility describes a system's susceptibility to an increase of its $n$-R\'enyi entropy in an interaction, it can be expected to be closely related to a system's susceptibility to receive quantum information, i.e., to receive pre-existing entanglement, as measured using $n$-R\'enyi entropies such as the von Neumann entropy, through an interaction. Follow-up work in this direction is in progress. 

Given its practical importance, a key type of quantum channel to apply the new methods to will be the quantum channels that arise through the light-matter interaction. Here, we already derived some results on the fragilities of both atom and field in a simple Unruh DeWitt model of the light-matter interaction. 

We finally note that our results on $n$-fragilities should relatively straightforwardly also apply in the field of quantum reference frames, see, e.g., \cite{Giacomini2019}. There, transformations from lab frames into the rest frame or into the center of mass frame of quantum particles are being considered. The corresponding transformations are generated by unitaries of the same form, $\exp(iA\otimes B)$, that we considered here. One of the observations regarding quantum reference frame transformations has been that coherence in one quantum reference frame can transform into entanglement in the transformed quantum reference frame. It should be very interesting to investigate this phenomenon using the new tool of  $n$-fragilities. 

$$$$
\bf Acknowledgements: \rm AK acknowledges support through a Discovery Grant from the National Science and Engineering Research Council of Canada (NSERC), a Discovery grant from the Australian Research Council (ARC) and through Google Faculty Research Awards.

\section*{References}
\bibliographystyle{unsrt}
\bibliography{iopart-num}

\begin{appendices}

\section{Qubit and ancilla qubit interacting through time evolution operator of the form $e^{it A\otimes H_B}$}
\label{appHb}

\begin{figure}
    \centering
    \includegraphics[trim={0 0 0 0},clip, scale=0.8]{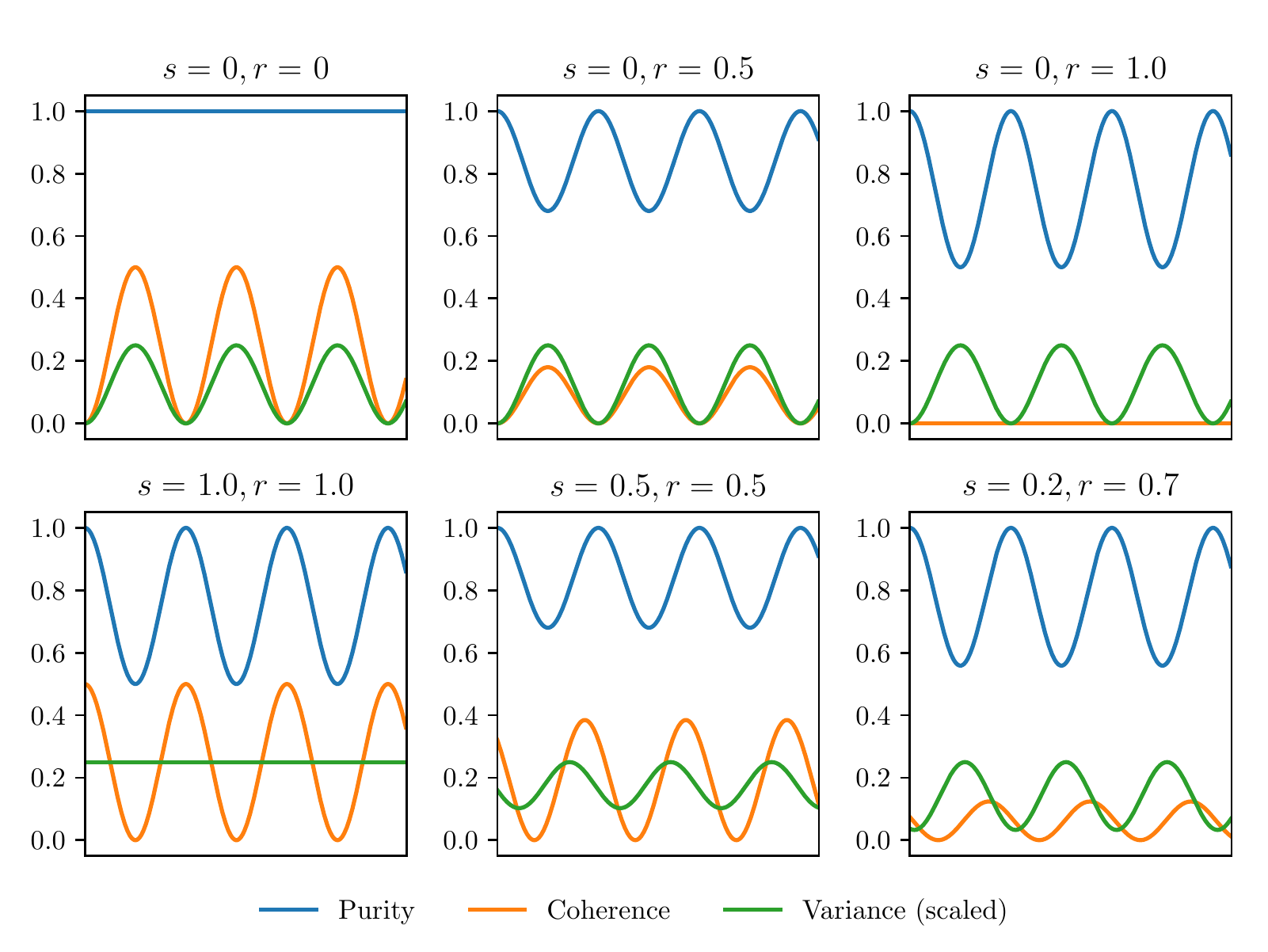}
    \caption{Time evolution of purity, 2-norm coherence under the action of Equation \ref{combu} for different values of $r$ and $s$ (chosen to be real for simplicity). The timescale (x-axis) can be chosen arbitrarily, and ($\vert b_x - b_y\vert$) has been normalised to 1 in order to plot the variance alongside the purity and coherence.}\label{u_p_c}
\end{figure}

For a qubit system and a qubit ancilla, we here consider a unitary time evolution operator of the form
\begin{equation}\label{combu}
    U(t)=e^{i\varepsilon t \sigma_x\otimes H_B},
\end{equation}
where $H_B$ does not commute with $B$ and is given by
\begin{equation}\label{hamb2}
    H_B=-i\big(\ket{b_x}\bra{b_y}-\ket{b_y}\bra{b_x}\big).
\end{equation}
If we apply Equation \ref{combu} to an initially pure, unentangled input state of the form given in Equation \ref{eq:psi}, then we find that the components of the reduced density matrix of the primary subsystem are given by (tracing out the ancilla)

\begin{align}
    &\rho_{xx} = \vert\chi^+\vert^2+\vert r\vert^2\vert\chi^-\vert^2,\\
    &\rho_{xy} = \rho_{yx}^* =  \chi^+\xi^{-*}+\vert r\vert^2\chi^-\xi^{+*},\\
    &\rho_{yy} = \vert\xi^-\vert^2+\vert r\vert^2\vert\xi^+\vert^2,
\end{align}
where we define: 
\begin{align}
    &\chi^+:= \frac{\cos(\varepsilon t)+s\sin(\varepsilon t)}{\sqrt{1+\vert r\vert^2}\sqrt{1+\vert s\vert^2}} \quad \chi^-:= \frac{\cos(\varepsilon t)-s\sin(\varepsilon t)}{\sqrt{1+\vert r\vert^2}\sqrt{1+\vert s\vert^2}}\nonumber\\
    &\xi^+:= \frac{s \cos(\varepsilon t)+\sin(\varepsilon t)}{\sqrt{1+\vert r\vert^2}\sqrt{1+\vert s\vert^2}} \quad \xi^-:= \frac{s \cos(\varepsilon t)-\sin(\varepsilon t)}{\sqrt{1+\vert r\vert^2}\sqrt{1+\vert s\vert^2}}\nonumber
\end{align}
The purity, 2-norm coherence, and uncertainty are then again defined through Equation \ref{eq:2d-purity}, Equation \ref{eq:2d-coherence}, and Equation \ref{eq:2d-uncertainty}, respectively. Figure \ref{u_p_c} demonstrates the relationship between these quantities for the primary qubit subsystem. 

Having chosen the interaction Hamiltonian of the form $\sigma_x\otimes H_B$, we expect that when the environment subsystem is initialised in an eigenstate of $\sigma_x$ the interaction will be non-entangling and the evolution of the primary subsystem will proceed unhindered under the Hamiltonian $\hat{H}_B$. Indeed, the top left subplot of Figure \ref{u_p_c} demonstrates this behaviour. The 2-norm coherence of the primary subsystem is initially zero, and grows with time to reach its maximal value of $1/2$. 
As we increase $r$ from zero, the initial state of the environment qubit moves away from an eigenstate of $\hat{\sigma}_x$, and the interaction becomes entangling. The top centre subplot of Figure \ref{u_p_c} demonstrates this effect. As entanglement is generated, the purity drops. This decreases the efficacy of the coherence generation process in the primary subsystem. Namely, the maximal coherence reached is lower than for the non-entangling interaction. 

However, an important feature to note is that the purity and coherence curves of this subplot are out of phase. That is to say, entanglement is being generated whilst the coherence of one subsystem is also increasing. The fact that these two processes are not necessarily anti-correlated highlights that it is important not to conflate the term "decoherence" (in the sense of a decrease in purity) with the changes in the 2-norm coherence itself. 

As $r$ is increased further, we reach the point at which the environment qubit state is maximally distinct from an eigenstate of $\hat{\sigma}_x$. For $r=1$, the initial state is instead an eigenstate of $\hat{\sigma}_z$. Here the interaction is maximally entangling, the initial value of the coherence in the primary subsystem constitutes an upper bound. This is demonstrated both in the top right subplot of Figure \ref{u_p_c} in which the coherence remains zero at all times, and in the bottom left subplot, in which the coherence is restricted to oscillate below the initial value. In both of these cases, it can be seen that the maximally entangled state is achieved (purity = 1/2). It is only possible to generate this maximally entangled state if $r=1$, as this corresponds to maximal coherence of the initial environment state, and since $\Delta \sigma_x$ does not change during the interaction (we assume no free evolution of either subsystem) the values of the coherence and the purity are restricted by Equation \ref{eq:2-dim-variance}.

While the effect of increasing $r$ from zero is to increase the entanglement-generating power of the interaction and decrease the efficacy of coherence generation, the effect of increasing $s$ from zero is to induce a phase shift between the purity and coherence curves. for $s=0$ these curves are 180 degrees out of phase (top centre and top right subplots), while for non-zero $s$ the phase difference decreases (bottom centre and bottom right). For $s=1$ the two curves are in phase, and because of this it is possible to cancel their contributions in equation Equation \ref{eq:2-dim-variance} such that the variance remains constant in time (bottom left subplot).

\begin{figure}
    \centering
    \includegraphics[trim={0 0 0 0},clip,scale=0.63]{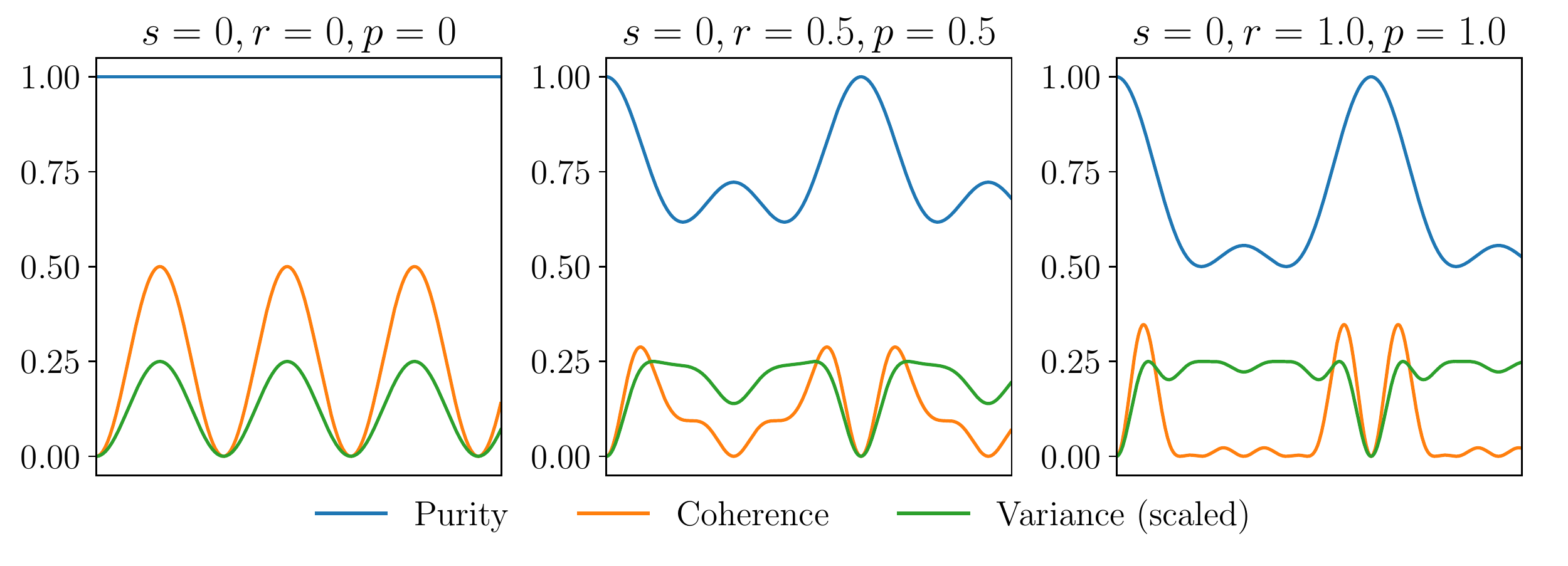}
    \caption{Examples of the time evolution of purity, 2-norm coherence, and variance of the primary subsystem of \ref{psi2} under the interaction \ref{combu2}, for sample values of $s,p,r$. The timescale (x-axis) can be chosen arbitrarily, and the spread of the eigenvalues has been normalised to 1 in order to plot the variance alongside the purity and coherence.}\label{u_p_c_2}
\end{figure}

A similar analysis to the above may be performed for cases where the environment system is more complicated than a qubit. For example, a superposition of Fock states under a unitary time evolution operator of the form
\begin{equation}\label{combu2}
    U(t)=e^{i\varepsilon t N\otimes H_B},
\end{equation}
where $N$ is the number operator, $a^\dagger a$, with eigenbasis consisting of the Fock states, $\{\ket{n}\}$. An example of an initial state we may then be interested in is 
\begin{equation}\label{psi2}
    \ket{\psi_0}=\frac{1}{\sqrt{\left(1+\vert r\vert^2+\vert p\vert^2\right)\left(1+\vert s\vert^2\right)}}\left(\ket{1}+r\ket{2}+p\ket{3}\right)\otimes\left(\ket{b_x}+s\ket{b_y}\right).
\end{equation}
While the time evolution of the primary subsystem is more complicated when coupled to such an environment, its purity, 2-norm coherence and variance are still related by Equation \ref{eq:2-dim-variance}. The degree to which the environment hinders the generation of coherence in the primary subsystem is dependent upon the parameters $p$ and $r$. Figure \ref{u_p_c_2} demonstrates this for a sample of possible values. 

\section{Time derivatives of the von Neumann entropy}\label{app:S_dot}
For a single system $\rho$, we have:
\begin{equation}\label{eq:S_2}
    \dot{S} = \displaystyle\lim_{\varepsilon \to 0} \frac{\dot{\gamma}_{1+\varepsilon}}{ \varepsilon}, 
    \qquad
    \ddot{S} = \displaystyle\lim_{\varepsilon \to 0}\frac{\ddot{\gamma}_{1+\varepsilon}}{\varepsilon},
\end{equation}
where 
\begin{equation}
    \gamma_n = \Tr \left[\rho^n\right], \qquad \dot{\gamma}_n = \Tr \left[\rho^{n-1}\dot{\rho}\right]
\end{equation}
We now consider the case $n = 1+ \varepsilon$ and the limit $\varepsilon \to 0$:
\begin{align}\label{lim_gamma}
    \displaystyle\lim_{\varepsilon \to 0} \dot{\gamma}_{1+\varepsilon} & = \Tr \left[\rho^\varepsilon \dot{\rho}\right]\nonumber \\
    & = \Tr\left[\exp(\varepsilon\ln (\rho))\dot{\rho}\right]\nonumber\\
    & = \Tr\left[(1+\varepsilon \ln (\rho))\dot{\rho}\right]\nonumber \\
    & = \varepsilon \Tr\left[\dot{\rho}\ln (\rho)\right]
\end{align}
We can then use the result \ref{lim_gamma} to calculate the time derivative of the von Neumann entropy:
\begin{equation}
    S = -\Tr \left[\rho \ln (\rho)\right]
\end{equation}
\begin{align}
    \therefore \ \dot{S} &= - \Tr \left[\dot{\rho} \ln (\rho)+\rho \  \partial_t \ln (\rho)\right] \nonumber \\
    & = -\displaystyle\lim_{\varepsilon\to 0}\frac{\dot{\gamma}_{1+\varepsilon}}{\varepsilon}-\Tr\left[\rho \ \partial_t\ln (\rho)\right]\nonumber \\
    & = -\displaystyle\lim_{\varepsilon\to 0}\frac{\dot{\gamma}_{1+\varepsilon}}{\varepsilon}-\displaystyle\lim_{\varepsilon\to 0}\Tr\left[\rho \ \partial_t\left(\frac{\rho^\varepsilon-\mathrm{I}}{\varepsilon}\right)\right]\nonumber \nonumber \\
    & = -\displaystyle\lim_{\varepsilon\to 0}\frac{\dot{\gamma}_{1+\varepsilon}}{\varepsilon}-\displaystyle\lim_{\varepsilon\to 0}\frac{1}{\varepsilon}\Tr\left[\rho \varepsilon\rho^{\varepsilon-1}\dot{\rho}\right]\nonumber \\
    & = -\displaystyle\lim_{\varepsilon\to 0}\frac{\dot{\gamma}_{1+\varepsilon}}{\varepsilon}-\displaystyle\lim_{\varepsilon\to 0}\Tr\left[\rho^{\varepsilon}\dot{\rho}\right] \nonumber \\
    & = -\displaystyle\lim_{\varepsilon\to 0}\frac{\dot{\gamma}_{1+\varepsilon}}{\varepsilon}-\displaystyle\lim_{\varepsilon\to 0}\Tr\left[(1+\varepsilon \ln (\rho))\dot{\rho}\right]\nonumber \\
    & = -\displaystyle\lim_{\varepsilon\to 0}\frac{\dot{\gamma}_{1+\varepsilon}}{\varepsilon}-\displaystyle\lim_{\varepsilon\to 0}\varepsilon\Tr\left[\dot{\rho}\ln (\rho)\right]\nonumber \\
    & = -\displaystyle\lim_{\varepsilon\to 0}\frac{\dot{\gamma}_{1+\varepsilon}}{\varepsilon}-\displaystyle\lim_{\varepsilon\to 0}\dot{\gamma}_{1+\varepsilon}\nonumber \\
    & = -\displaystyle\lim_{\varepsilon\to 0}\frac{\dot{\gamma}_{1+\varepsilon}}{\varepsilon}
\end{align}

We can proceed in the same way to obtain the second time derivative as given above.



\section{Second time derivative of the n-purity}\label{app:gamma_ddot}
For a single system $\rho$, the second time derivative of the n-purity is given by:
\begin{align}\label{gamma_t2_result}
    \ddot{\gamma}_n = n \Tr\left[\sum_{i = 0}^{n-2}-\rho^i[H,\rho]\rho^{n-2-i}[H,\rho]+2\rho^{n-1}H\rho H -2 \rho^n H^2\right]
\end{align}
To see this, we first consider $\dot{\gamma}_n$:
\begin{align}
    \dot{\gamma}_n &= \partial_t\Tr[\rho^n]\nonumber \\
    & = n\Tr[\rho^{n-1}\dot{\rho}]\nonumber \\
    \therefore \ddot\gamma_{n} & = n \Tr\left[\sum_{i=0}^{n-2}\rho^i\dot{\rho}\rho^{n-2-i}\dot{\rho}+\rho^{n-1}\ddot{\rho}\right]\label{gamma_t2},
\end{align}
where the sum arises due to the assumed non-commutativity of $\rho$ and $\dot{\rho}$.
Substituting $\dot{\rho}=-i[H,\rho]$ and $\ddot{\rho} = -H^2\rho - \rho H^2+2H\rho H$ into equation \ref{gamma_t2} we obtain the expression \ref{gamma_t2_result}. We note that for a unipartite system, Equation \ref{gamma_t2_result} gives zero. To see this, we first expand the summation as follows:
\begin{align}\label{eq:expanded_sum}
    &-\Tr\bigg[\sum_{i=0}^{n-2}\rho^i[H,\rho]\rho^{n-2-i}[H,\rho]\bigg]\nonumber\\
    &=-\Tr\bigg[\sum_{i=0}^{n-2}\underbrace{2\rho^{i+1}H\rho^{n-1-i}H}_{1^{st}}-\underbrace{\rho^i H\rho^{n-i}H}_{2^{nd}}-\underbrace{\rho^{i+2}H\rho^{n-2-i}H}_{3^{rd}}\bigg],
\end{align}
where the 1$^{st}$, 2$^{nd}$, and 3$^{rd}$ components of the $i^{th}$ term have been indicated. With this labelling in mind, it is clear that
the 1$^{st}$ component of the $i^{th}$ term will be exactly cancelled by the 2$^{nd}$ component of the $i+1^{th}$ term and the 3$^{rd}$ component of the $i-1^{th}$ term. Cancelling in this way will leave behind the 1$^{st}$ and 2$^{nd}$ components of the $i=0$ term, the 2$^{nd}$ component of the $i=1$ term, the 3$^{rd}$ component of the $i=n-3$ term, and the 1$^{st}$ and 3$^{rd}$ components of the final $i=n-2$ term. That is to say, the sum collapses to:
\begin{equation}
    2\rho H\rho^{n-1}-H\rho^n H-\rho H\rho^{n-1} H - \rho^{n-1}H\rho H+2\rho^{n-1} H\rho H -\rho^n H^2.
\end{equation}
Now employing the cyclicity of the trace, we see that Equation \ref{eq:expanded_sum} becomes
\begin{equation}
    \Tr[-2\rho^{n-1}H\rho H+2\rho^n H^2],
\end{equation}
which exactly cancels with the final two terms outside the summation in Expression \ref{gamma_t2_result}.

\section{Further results for the light-matter interaction toy model}\label{app:lm-plots}

\begin{figure}[H]
\centering
\begin{tabular}{cc}
{\includegraphics[scale = 0.55, trim={1.5cm 0cm 0cm 0cm}]{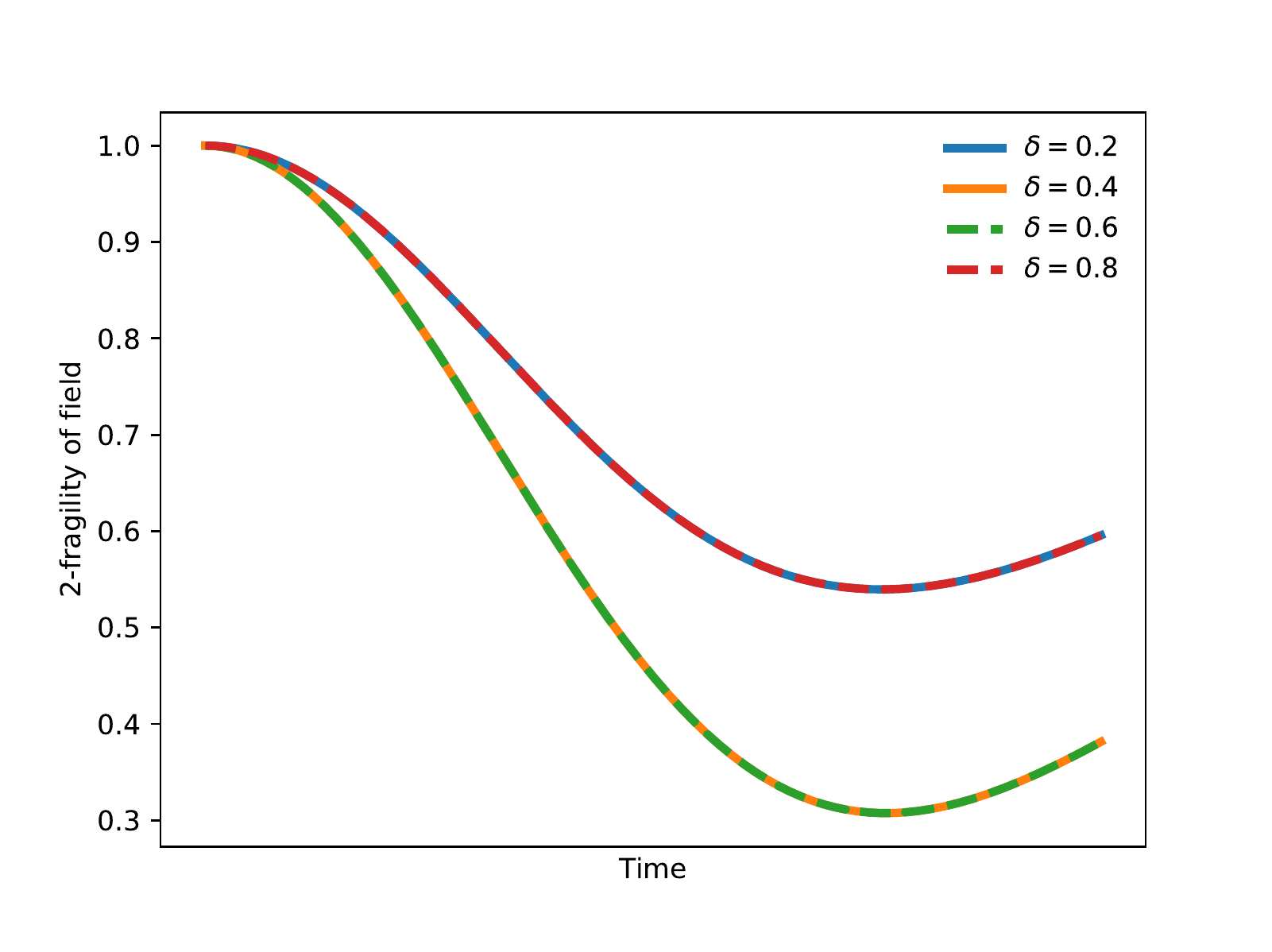}} &
{\includegraphics[scale = 0.55, trim={1.5cm 0cm 0cm 0cm}]{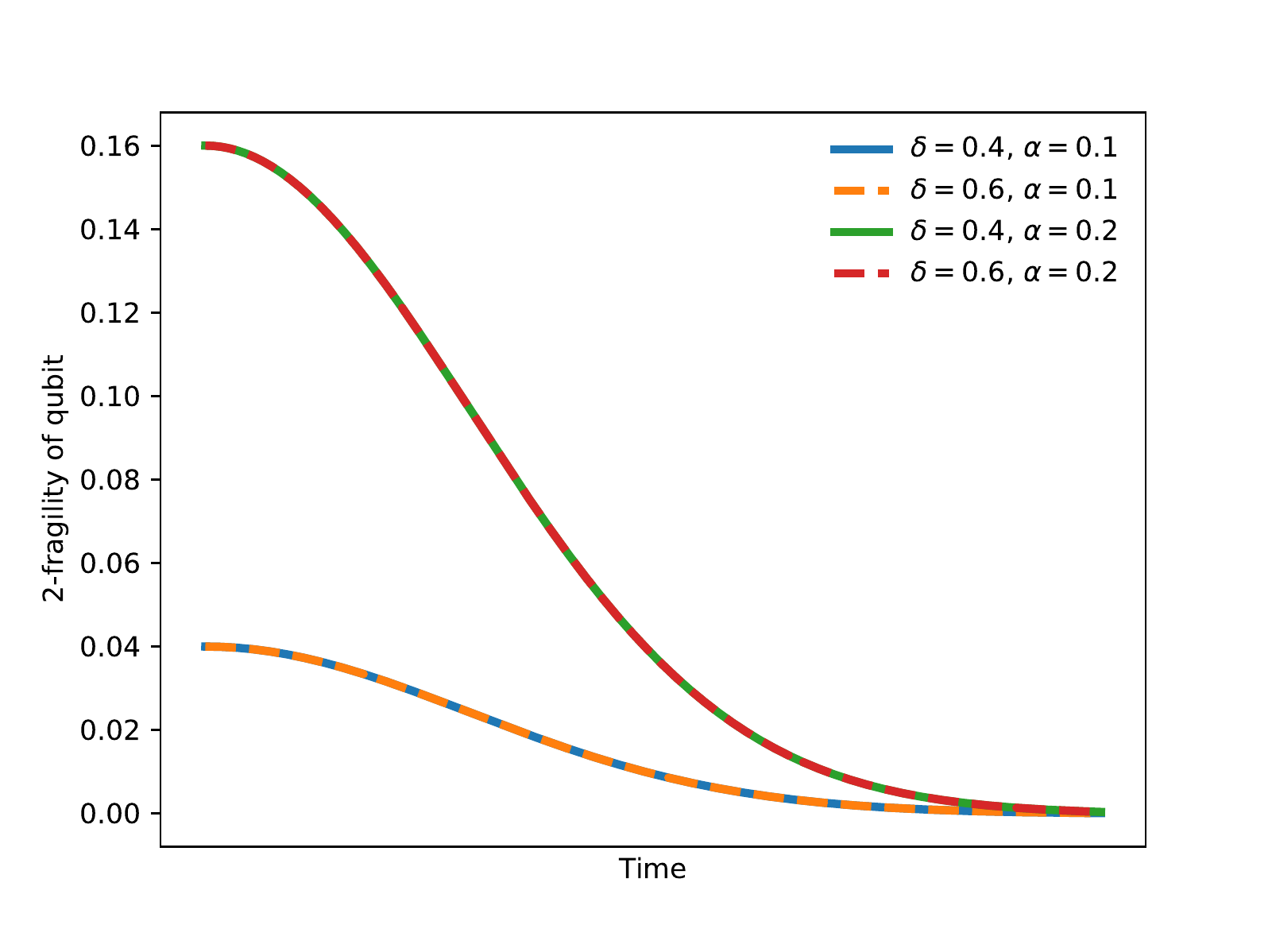}}
\end{tabular}
\caption{Comparisons of the evolution of the 2-fragility for the field (left) and the qubit (right), for different initial conditions. While the fragility of the field depends only on $\delta$, the fragility of the qubit instead depends only on $\alpha$.}\label{fig:fragilities-comp}
\end{figure}

\begin{figure}[H]
\centering
\begin{tabular}{cc}
{\includegraphics[scale = 0.55, trim={1.5cm 0cm 0cm 0cm}]{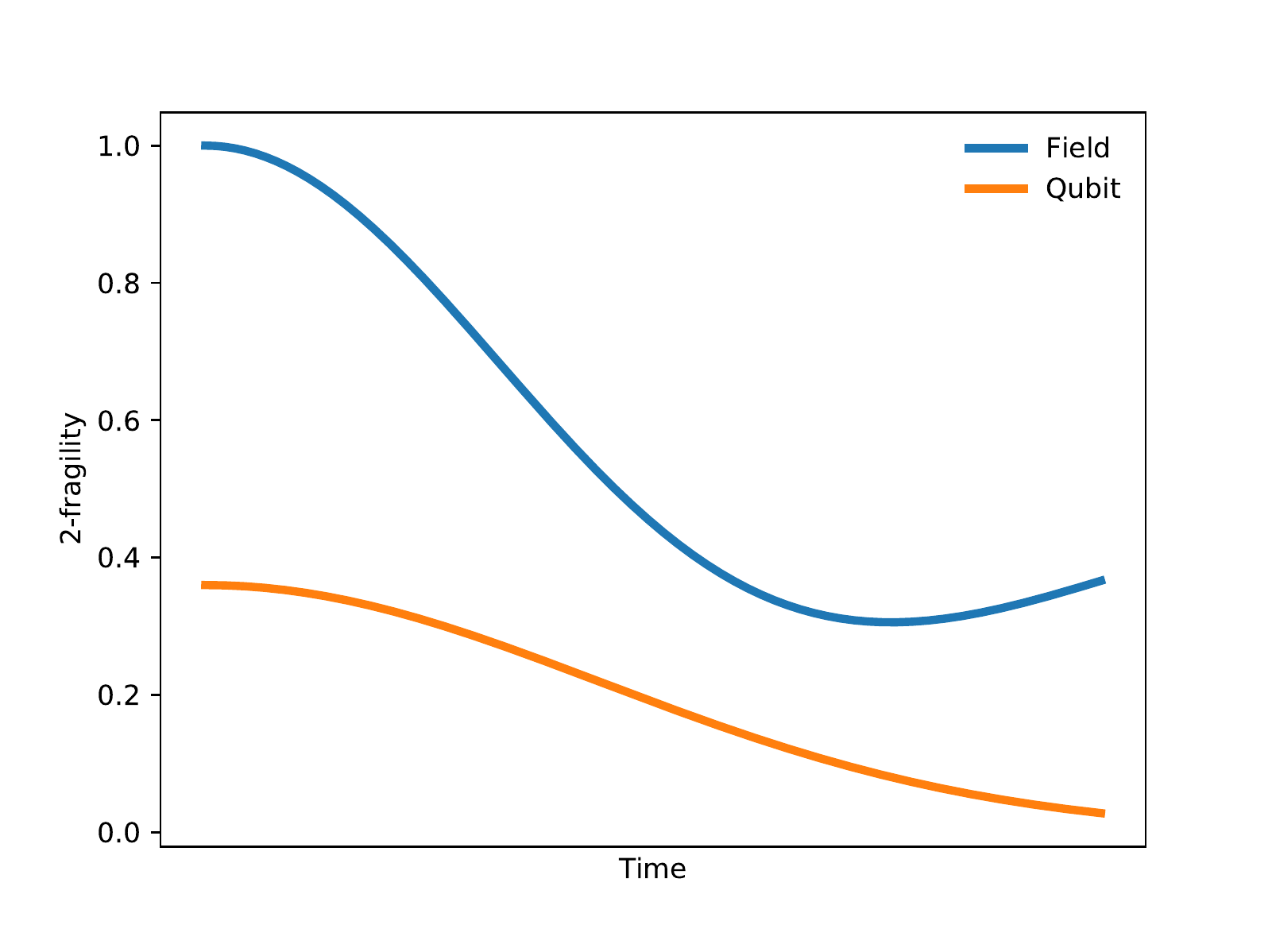}} &
{\includegraphics[scale = 0.55, trim={1.5cm 0cm 0cm 0cm}]{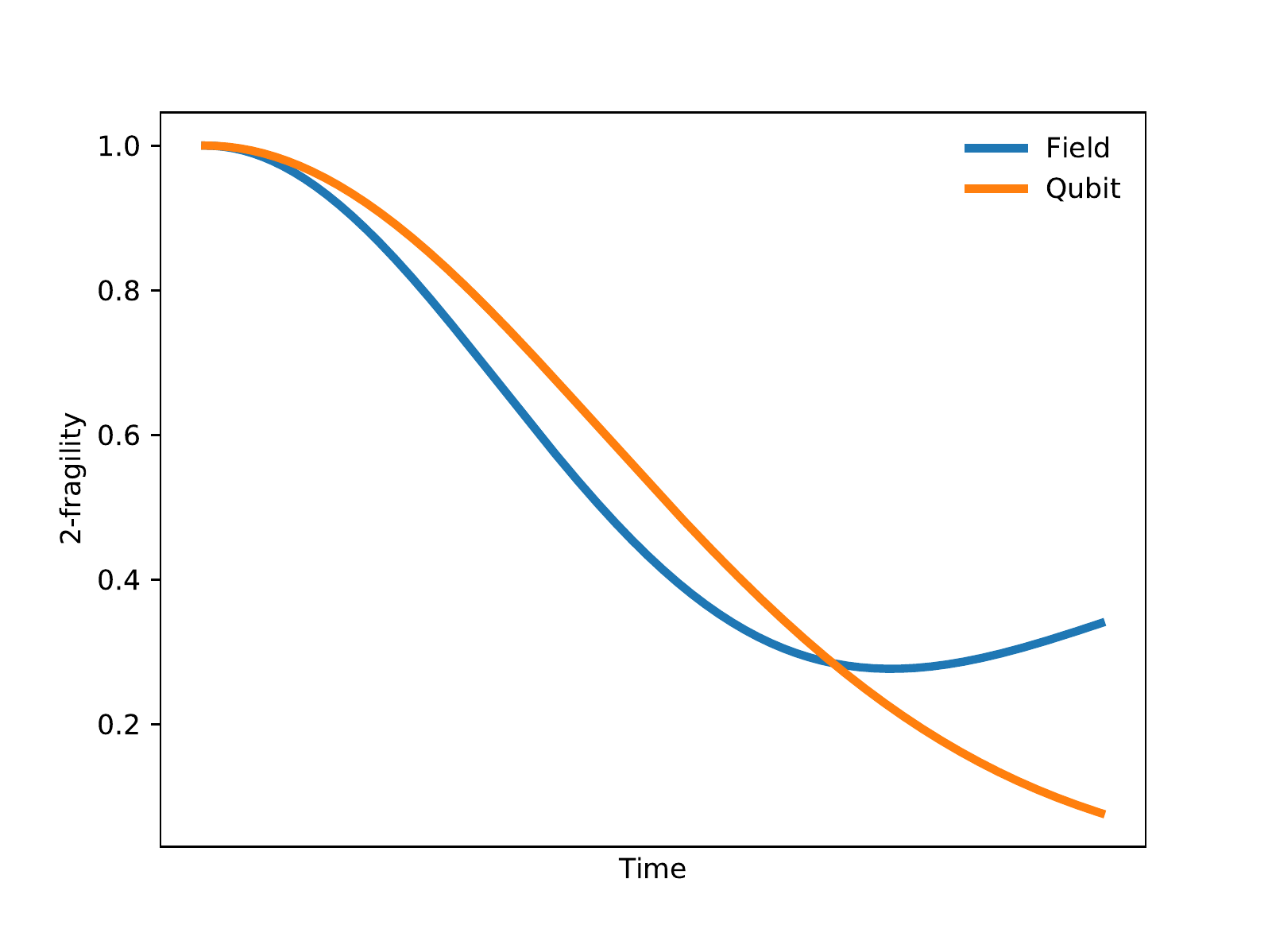}}
\end{tabular}
\caption{Fragilities of both the field and the qubit plotted alongside one another for different initial states. The left plot represents $\delta=0.4$, $\alpha=0.3$, while the right plot represents $\delta=0.5$, $\alpha=0.5$}\label{fig:fragilities-both}
\end{figure}

\begin{figure}[H]
\centering
\begin{tabular}{cc}
{\includegraphics[scale = 0.55, trim={1.5cm 0cm 0cm 0cm}]{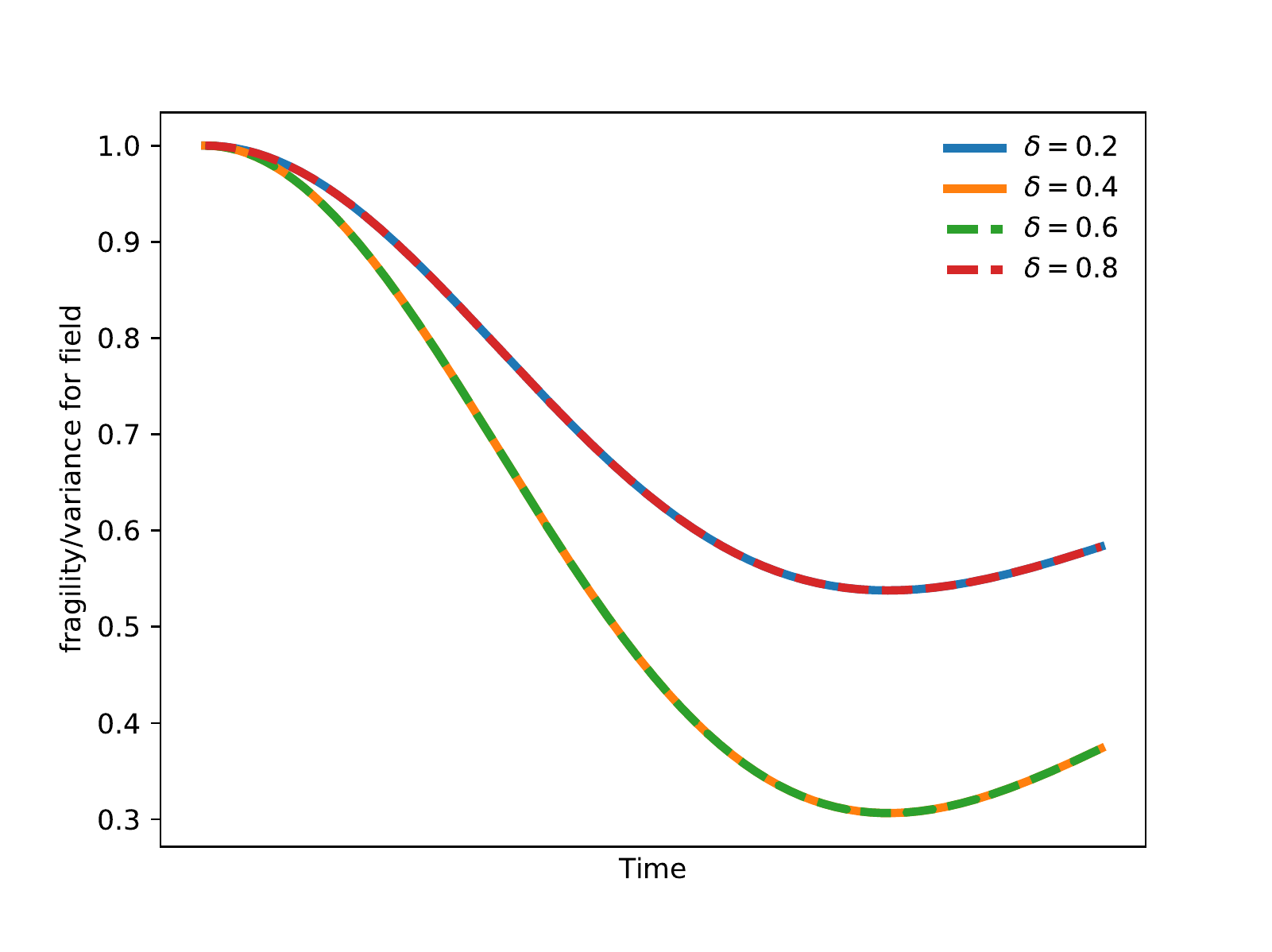}} &
{\includegraphics[scale = 0.55, trim={1.5cm 0cm 0cm 0cm}]{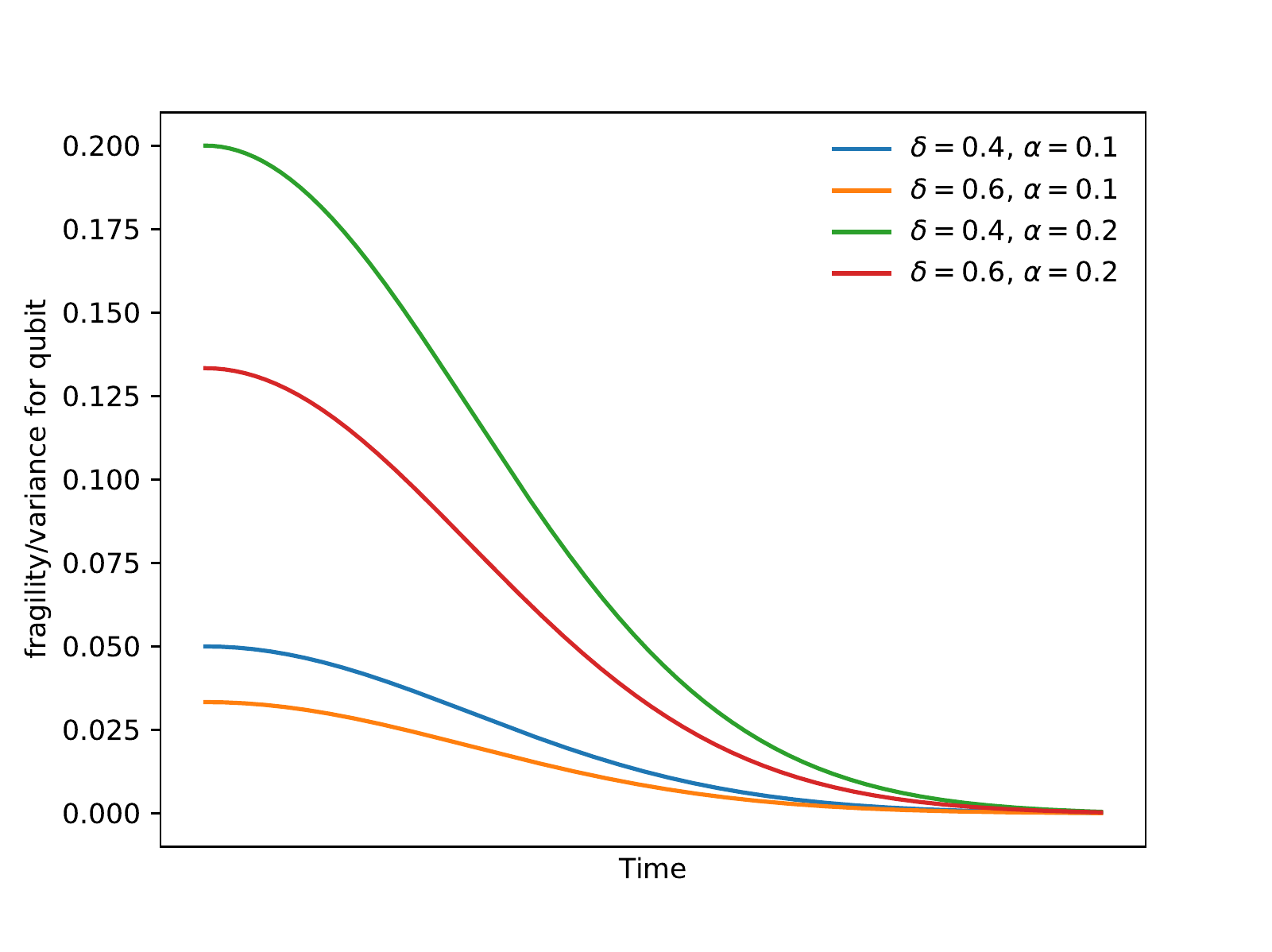}}
\end{tabular}
\caption{Time evolution of the ratio of the 2-fragility to the variance for the field (left) and the qubit (right). In both cases the variance is constant, however because the variance depends on $\delta$ in the qubit case while the fragility depends on $\alpha$, we see a separation of the curves in both parameters. For the field, neither the fragility nor the variance depend on $\alpha$.}\label{fig:fragilities-again}
\end{figure}

\end{appendices}

\end{document}